\documentclass[]{mnras}
\usepackage{amsmath}
\usepackage{graphicx}
\usepackage[T1]{fontenc}
\usepackage{aecompl}

\title[A radiation transfer model for the radiation-fields in the Milky-Way]
{A radiation transfer model for the Milky-Way: I. Radiation fields and 
application to High Energy Astrophysics}
\author[C.C. Popescu et al.]{C. C. Popescu$^{1,2,3}$\thanks{E-mail:
cpopescu@uclan.ac.uk}, R. Yang$^{3,4}$, R.J. Tuffs$^{3}$, 
G. Natale$^{1}$, M. Rushton$^{2}$, F. Aharonian$^{3,5}$\\
$^{1}$ Jeremiah Horrocks Institute, University of Central Lancashire, PR1 2HE, 
Preston, UK\\
$^{2}$ The Astronomical Institute of the Romanian Academy, Str. Cutitul de
Argint 5, Bucharest, Romania\\
$^{3}$ Max Planck Institut f\"ur Kernphysik, Saupfercheckweg 1, 69117
Heidelberg, Germany\\
$^{4}$ Key Laboratory of Dark Matter and Space Astronomy, Purple Mountain
Observatory, Chinese Academy of Sciences, \\Nanjing, 210008, China\\
$^{5}$ School of Cosmic Physics, Dublin Institute for Advanced Studies, 31 
Fitzwilliam Place, Dublin 2, Ireland}
\begin{document}

\date{Accepted . Received ; in original form }

\pagerange{\pageref{firstpage}--\pageref{lastpage}} \pubyear{2002}

\maketitle

\label{firstpage}

\begin{abstract}
We present a solution for the ultraviolet (UV) - submillimeter (submm) 
interstellar radiation fields (ISRFs) of the Milky Way, derived from modelling
COBE, IRAS and Planck maps of the all-sky emission in the near-, mid-, far-infrared and submm.
The analysis uses the axisymmetric radiative transfer (RT) model 
that we have previously implemented to model the panchromatic spectral
energy distributions (SEDs) of star forming galaxies in the nearby
universe, but with a new methodology allowing for optimisation of
the radial and vertical geometry of stellar emissivity and dust opacity,
as deduced from the highly resolved emission
seen from the vantage point of the Sun. As such, this is the first self-consistent
model of the broad-band continuum emission from the Milky Way.
In this paper, we present model predictions for the spatially integrated
SED of the Milky Way as seen from the Sun, showing good agreement with the data,
and give a detailed description of the
solutions for the distribution of ISRFs, as well as their physical origin,
throughout the volume of the galaxy. We explore 
how the spatial and spectral distribution of our new predictions for the ISRF 
in the Milky Way affects the amplitude and spectral distribution of the 
gamma-rays produced via Inverse Compton  scattering for cosmic ray electrons situated
at different positions in the galaxy, as well as the 
attenuation of the gamma-rays due to interactions of the gamma-ray photons 
with photons of the ISRF.  We also compare and contrast our solutions for the
ISRF with those incorporated in the GALPROP
package used for modelling the high energy emission from cosmic rays in the Milky Way.

\end{abstract}


\section{Introduction}

Cosmic rays (CR) are a major energetic constituent of the interstellar medium 
(ISM) of the Milky Way and other galaxies, contributing a comparable energy 
density to that in photons, magnetic field and thermal gas. They control key 
processes governing galactic evolution both on large scales, such as powering 
galactic winds, and on small scales, such as in dense star-forming cores of 
molecular clouds, where they are the dominant ionising agent enabling magnetic 
pressure support of the cores. Despite this, the questions of what the global
distribution of CRs is within the galaxy, where they originate, and how they propagate in the ISM,
are still open. 
One fundamental reason for these uncertainties is that 
the solar neighbourhood is still the only place in the Galaxy where we have a 
truly unambiguous measurement of the amplitude and spectrum of the flux of CR 
protons and electrons. At any other position, CRs are usually detected via proxies, through gamma-ray emission in the MeV to TeV spectral range, which has two major components: i) gamma-rays arising from the 
decay of pions produced in the interaction of CR protons and nucleons with 
interstellar neutral and molecular hydrogen (the hadronic component); 
ii) gamma-rays arising from the Inverse Compton scattering by CR 
electrons\footnote{CR electrons in the ISM can also be detected in gamma-rays via
relativistic bremsstrahlung emission, and at radio wavelengths through
synchrotron radiation} and
positrons of photons in the cosmic microwave background (CMB) radiation field 
and the interstellar radiation fields (ISRF) (the leptonic component).
 Because the contribution from these two main competing mechanisms is in many cases
 comparable and their spectral shape not markedly dissimilar (Aharonian \&
 Atoyan 2000), 
extracting robust information on CR fluxes based on the overall level of 
gamma-ray emission and on its spectral shape has proven to be difficult. 
Therefore, additional constraints need to be 
considered, in particular through prior knowledge of gas column in the 
case of pion decay, and of ISRFs in the case of the Inverse Compton effect. 
While
radiation fields from the CMB are important in providing seed
photons for the very energetic TeV gamma-ray photons, it is the
10 to 1000\,${\mu}$m mid-infrared (MIR), far-infrared (FIR) and submillimetre 
(submm) radiation  fields, originating from dust emission, that provide the 
seed photons for an important part of the gamma-ray spectrum in the GeV-TeV 
range, and the 0.1-10\,${\mu}$m ultraviolet (UV), optical and near-infrared (NIR) 
radiation fields originating from stellar sources that power the MeV 
gamma-ray emission. In addition, the infrared radiation fields affect
the propagation of gamma-rays through the intergalactic and interstellar 
medium (Moskalenko et al. 2006) through pair production of gamma-ray photons 
on the background radiation fields. 

That there has been relatively little focus on the quantification of ISRFs in galaxies in general,
and in the Milky Way in particular, may be due to
the fact that they are not
directly measurable and thus difficult to determine. It would be 
relatively straightforward to determine 
radiation fields by direct observations of the sources of photons, which, in
most cases, are predominantly stars, if galaxies were optically thin.
However, galaxies - and in particular star-forming galaxies - contain
dust grains, which, because they absorb and scatter photons, partially
or wholly prevent a direct measurement of the spatial and spectral
distribution of the sources of the RFs in the UV/optical. The propagation of light depends
in a complex way on the relative distribution of stellar emissivity
and dust opacity, with structures ranging in scale from parsecs
to kiloparsecs.  However,  the stellar light that is absorbed by dust is 
re-radiated in the infrared/submm range, with spectral characteristics directly
related to the heating by stellar radiation fields. Thus, images
of the amplitude and colour of the infrared/submm emission can potentially place 
strong constraints on the distribution of the UV/optical ISRF within galaxies,
even when stellar populations are obscured. Moreover, if, as is almost always the case,
galaxies are optically thin in the far-infrared/submm, observed images 
can provide a more direct constraint on the spatial distribution of the emissivity, and thereby the  
ISRF, in these bands.

Work on modelling external galaxies observed in the infrared/submm has confirmed that 
radiative transfer techniques can be used to self-consistently link the 
radiation fields to the spectral energy distributions (SEDs) in the
UV/optical/infrared/submm, in combination with independent empirical 
constraints on the spatial distribution of young and old stellar populations 
and of dust (Popescu et al. 2000a, Misiriotis et al. 2001, Popescu et al. 2004, 
Bianchi 2007, Baes et al. 2010, Bianchi \& Xilouris 2011, MacLachlan et
al. 2011, De Looze et al. 2012a,b, 2014). 

Since the Milky Way is the best observed galaxy, and our nearest astrophysical laboratory,
it is clearly desirable to extend the self-consistent RT analysis of the
panchromatic emission of external spiral galaxies to determine the UV/optical/infrared/submm
radiation fields throughout the volume of the interstellar medium of our own galaxy.  
Here we present results of such an analysis,
obtained using a development of the axisymmetric RT model that 
we have previously implemented to model the integrated panchromatic SEDs of 
nearby spiral galaxies  (Popescu et al. 2000a, Tuffs et al. 2004, 
Popescu, Tuffs et al. 2011, referred to as PT11 in this paper).
This model has been successful in accounting for both the spatially integrated SEDs of 
individual galaxies (Popescu et al. 2000a, Misiriotis et al. 2001) and to
predict the appearance of the modelled galaxies in the dust emission (Popescu
et al. 2004, Popescu et al. 2011). Generic solutions for the distribution
of ISRF in external galaxies implicit to the model SEDs presented in PT11
are given in Popescu \& Tuffs (2013). The PT11 model has also predicted the 
statistical behaviour of a variety of observables of the population of
spiral galaxies in the local Universe (Driver et al. 2007, 2008, 2012;
Graham \& Worley 2008; Masters et al. 2010; Gunawardhana et al. 2011;
Kelvin et al.  2012, 2014, Grootes 2013a;  Pastrav et al. 2013a,b; Vulcani et
al. 2014; Devour \& Bell 2016). 
Grootes et al. (2013b) has shown that using the RT model of PT11 to correct the
fundamental scaling relation between specific star-formation rates, as measured from the UV continuum, and stellar mass for the effects of dust attenuation leads to a marked reduction of the scatter in this relation, confirming the ability of the PT11 model to predict the propagation 
of UV light in galaxies. Recently Davies et al. (2016) has shown that, when critically compared
and contrasted with various methods to derive star-formation rates in
galaxies, the one using this RT method gives the most 
consistent slopes and normalisations of the specific SFR-stellar mass 
relations. 

Previous studies of the large-scale distribution of stars and diffuse
dust emission in the Milky-Way have often separately modelled either
star count data or the dust emission, without linking the two by a RT calculation to
explicitly predict the infrared/submm emission of grains in response to the ISRF at each
position in the galaxy.
Stellar distribution models include the Besan\c con model (Robin \& Creze 1986; 
Bienayme et al. 1987; Robin et al. 1996, 2003), the SKY model
(Wainscoat et al. 1992; Cohen 1993, 1994, 1995), and the TRILEGAL
model (Girardi et al. 2005). Models of the diffuse dust emission in
the Milky Way have also been developed to reproduce the large-scale
mid- and far-infrared emission: Sodroski et al. (1997),  Davies et
al. (1997), Drimmel (2000),  Drimmel \& Spergel
(2001) and  Misiriotis et al. (2006). In the cases that self-consistent RT
models have been employed, these have  been used to only model a very
  limited wavelength and spatial range, or have not been optimised to fit
  all-sky emission observational data. Thus, Robitaille et al. (2012) developed
  a non-axisymmetric RT model
  of the Milky Way which was only compared to the mid-infrared data and only
  for 274 deg$^2$. Because only a very narrow strip in latitude was used for
  comparison purposes, no handle on the contamination from highly resolved 
foreground emission from structures local to the Sun could have been applied.
 In addition the model does not incorporate local 
absorption and emission in the star forming regions, which are the main 
contributors to the mid-infrared emission in star forming galaxies in the 25 
and 60 micron bands.
 Another RT model is that incorporated in the GALPROP
package, developed and continuously improved in
the last 16 years (Strong et al. 2000; Moskalenko et al. 2002; 
Porter \& Strong 2006; Moskalenko et al. 2006; Porter et al. 2008). This is an 
axisymmetric model providing an explicit
calculation of the radiation fields in the Milky Way,
which was specifically designed for use in studies of the 
high energy gamma-ray emission. However this model was not optimised to fit
the all-sky observational data in the NIR-FIR-submm, but was built using the 
assumption that the stellar and dust distributions are known from stellar counts
and gas measurements. As with the Robitaille model, the GALPROP model does not 
incorporate the contribution of star forming regions, and was recently 
shown to not predict the observed IRAS, COBE and Planck maps of the Milky Way 
(Porter 2016).

In this context it is to be noted that
modelling the panchromatic emission from the Milky Way is in practice a 
far more challenging problem than for external galaxies. This is in large measure due to the
position of the Sun within the galaxy, which induces a massive degeneracy, not
present in the extragalactic studies, 
between the distance to the observer and the luminosity of emitters. This
difficulty is compounded by the fact that, from the vantage point of the Sun,
dust in the galactic plane obscures almost all direct stellar light from the inner galaxy 
shortwards of 3 micron; by contrast, even in an edge-on view,
visual measurements can still be used to constrain the RT analysis of external spiral galaxies.
The luminosity-distance degeneracy may potentially be sidestepped by assuming 
some physical link between dust grains and gas in galaxies,
since radio spectroscopic observations of gas tracers (primarily the hyperfine
21cm Hydrogen line to trace HI and rotational transitions of the CO molecule to trace H$_{2}$)
can be used to model the distribution
of the gas. Indeed, this approach has been commonly adopted in modelling imaging observations
of the dust emission, including the RT analyses of Porter et al. (2008) and Robitaille et al. (2012).
However, the transformations between the radio tracers and the actual distribution of the
gas are themselves challenging to physically model and are empirically uncertain. This,
in turn, may potentially introduce systematic error into the model predictions for the ISRF. Moreover
(and motivated by the empirical uncertainty in the conversion factors of the radio tracers), one 
fundamental goal of
modelling the gamma-ray emission from the Milky Way will ultimately be to use the 
pion-decay component  of the emission to determine the hydrogen gas distribution in the Milky Way
completely independently of the radio tracers; for such an analysis one would also
require that the predictions of the inverse Compton component of the gamma-ray emission
be independent of the radio tracers.

Motivated by these considerations, the observational input for our RT model for the panchromatic
emission from the Milky Way is restricted to the distributions of surface brightness of the
galaxy provided by the COBE, IRAS and Planck maps of the all-sky emission in the near-, mid-, far-infrared and submm. In a separate paper (Natale et al. 2017) we describe in detail the modifications 
to the methodology of the PT11 model to allow the radial
and vertical distributions of stellar emissivity and dust opacity to be deduced from the highly resolved
emission maps.  
In this approach we are in effect using an assumption 
of axisymmetry to break the luminosity-distance degeneracy, and related 
degeneracies affecting the radial distributions of dust and stellar 
emissivity. 

In this paper, we present the 
model predictions for the corresponding 
solutions for the spatial and spectral dependencies of the solutions for the ISRFs from the UV to the submm. We also make these solutions available in 
electronic format for use in the analysis of gamma-ray emission in our Galaxy. 
These should help interpret the 
broad range of gamma-ray  observations that have been recently carried with 
the Fermi Gamma-ray Space Telescope in the energy range from 100 MeV to 300 
GeV and with the High Energy Stereoscopic System (HESS) ground-based Cherenkov 
telescope at the high-energy end in the range from 0.1 TeV to 10 TeV. 
This will constrain the interpretation of the large-scale diffuse gamma-ray emission of the 
Milky Way, in effect by imposing consistency between the inverse-compton component
of the gamma-ray emission and the completely independent
NIR-submm data on direct and dust-reradiated starlight. It will also constrain the
interpretation of discrete gamma-ray sources embedded in 
the diffuse ISM, such as plerions, shell-type SNRs, binary systems, or
other as yet unidentified sources of gamma-rays. 

The structure of the paper is as follows: In
Sect.~\ref{sec:model} we give an overview of the RT model used for the
calculations of the radiation fields in the Milky Way. In
Sect. ~\ref{sec:comp} we show the predictions of our model for
the dust emission SED of the Milky Way and compare this with the
observed spatially integrated SED derived from IRAS, COBE and Planck maps. We also show predictions for the local ISRF extending into the UV/optical regime and compare this with observational
constraints on the local UV/optical ISRF. The results of our model are
described in terms of radial and vertical profiles and SEDs of radiation fields
in Sect.~\ref{sec:results}. The resulting radiation fields presented
in this paper are compared with those predicted by the GALPROP
model in Sect.~\ref{sec:comp_galprop}. In Sect.~\ref{sec:HE} we
explore how the spatial and spectral distribution of the ISRF affects the 
spatial and spectral distributions of the
gamma-rays produced via Inverse Compton scattering, as well as the
attenuation of the gamma-rays due to interactions of
the gamma-ray photons with photons of the 
ISRF. In Sect.~\ref{sec:accelerators} we discuss the influence of the
modelled ISRF on predicted VHE Emission from accelerators of CR
electrons. In Sect.~\ref{sec:summary} we give a
summary of our main results.

\section{The model}
\label{sec:model}

Our model is based on the axisymmetric RT model of PT11 for the UV to submm emission
of external galaxies, in which
the geometry of dust opacity and stellar emissivity is prescribed by
parameterised functional forms. 
In the PT11 model, values of all geometric parameters 
had fixed ratios to the scale length of an exponential disk describing stellar
emissivity in the B-band,
with the ratios being calibrated either from  resolved observations of translucent
components of edge-on spiral galaxies
exhibiting dust lanes (Xilouris et al. 1999), or, for highly obscured structure
associated with young stellar populations, set according to physical considerations,
and such that the observed amplitude and spectrum of the observed
integrated infrared emission from dust could be predicted.
A key ingredient of the model for external galaxies, which we also
found necessary to reproduce the observed images of
the Milky Way in both direct and dust-reradiated starlight,
is the use of separate
disks to describe the emissivities of the young and old stellar
populations, with the latter having a larger scale height
than the former. As for the external galaxies, we also
needed to invoke separate geometrically thin and thick dust disks to be respectively associated with the
disks of young and old stellar emissivity in the Milky Way.

While retaining the overall formalism of the PT11 model as applied 
to the Milky Way, we nevertheless had to implement a new methodology that 
deals with the inner view of a galaxy and with the lack of direct 
observational constraints in the UV-optical regime within the solar circle.
We thus optimised the geometrical
components of the model and the value of the model parameters to fit
the observed maps seen from the position of the Sun in the submm, FIR
and NIR. The description of this detailed modelling is given in Natale et
al. (in prep). Here we only mention its main characteristics that are relevant for the 
understanding of the solution for the radiation fields. 
The main concept in the optimisation procedure is that we made use of the fact that each waveband is strongly affected only by a subset of parameters. 
   For example, the observed 
latitude and longitude profiles in the submm provided strong constraints for 
the radial and vertical distribution of diffuse dust. The distribution of 
24\,${\mu}m$ 
emission constrained the distribution of locally-obscured star forming regions, while the NIR 
emission was used to derive the characteristics of the old stellar 
populations, including their optical emission. The maps at the peak of the 
dust emission SED were important in deriving the distribution of diffuse UV 
heating, otherwise not directly accessible to observations. 
Therefore we were able to compartmentalise the optimisation problem and
introduce a hierarchy into this which avoided the need to simulataneously 
solve for all geometrical parameters. After solving for each subdomain in the hierachy of parameters we revisited the other domains to ensure a fully self-consistent 
calculation over all parameters. Several cycles over this hierarchy were needed
to converge on  the final solution. In each case the best fitting model
was determined by visually comparing the predicted
brightness profiles in longitude averaged
over latitude, and in latitude averaged over longitude
with the corresponding observed brightness profiles.
When needed (in particular to verify the model
prescription for the variation of scale height with radius), a comparison
was also made between model and data for strip latitude
profiles at various longitudes. In the context of the applications 
to gamma-ray astronomy, we found that the
predicted amplitudes of the infrared component of the ISRF, which is
the most important for predicting inverse compton emission from CR electrons, 
were especially sensitive to the vertical scale of the dust distribution. 

The result of this 
new optimisation technique led to a solution for the distribution of dust and 
stars that were able to self-consistently reproduce the overall average 
latitude and longitude profiles derived from the COBE maps at 1.2, 2.2, 3.5, 4.6, 140, and 240\,${\mu}m$,
the IRAS maps at 24, 60, and 100\,${\mu}m$, and the Planck maps at
350, 550 and 850\,${\mu}m$ (see Natale et al. in prep). Good agreement 
was obtained between model and data at all wavelengths except at 1.2${\mu}m$, 
where the model underestimates the emission by $27\%$. The origin of this
discrepancy is discussed in Sect.~\ref{subsec:limitations} and will be 
addressed in future work. Thus the optimisation was done using all available 
bands except the 12\,${\mu}m$ IRAS band. This band was not used as the
12\,${\mu}m$ flux from the Milky Way is dominated by emission from Polycyclic 
Aromatic Hydrocarbons (PAH) in the diffuse ISM, and the abundance of PAH, 
though well calibrated at the solar circle, might be expected to vary with 
galactocentric radius,  as is observed in external spiral galaxies. We
used the IRIS maps generated from IRAS data by Miville-Desch\^enes \& Lagache
(2005), the COBE/DIRBE maps from the CADE\footnote{Centre d'Analyse de
Donn\'ees Etendues (CADE; Analysis Center for extended data)
provides the COBE maps in the HealPix sky pixelisation scheme at
http://cade.irap.omp.eu/dokuwiki/doku.php?id=cobe} and Planck maps
obtained with the Planck High Frequency Instrument 
(HFI) as described in Ade et al. (2014) and Planck Collaboration VIII (2015). 

To avoid the optimisation being overly influenced by nearby emission structures,
both the model and observed maps were background-subtracted, using a background reference
determined from strips offset from the galactic plane by $\pm$~5 degrees in latitude. This
also filters out any non-local structure on large scales due to a galactic halo. Since, in any case,
it is difficult to distinguish between the contributions of distant halo emission and local
emission from the galactic plane above the Sun, all predictions for the ISRF given in this
paper are confined to the contributions from the galactic disk and bulge
only.

The RT calculations were made using both a modified version of the 2D 
ray-tracing radiative transfer code of Kylafis \& Bahcall (1987), as well as
the 3D ray-tracing RT code DART-Ray (Natale et al. 2014, 2015). The 2D 
code uses the method of scattered intensities, introduced by Kylafis \& Bahcall
(1987) in an implementation by Popescu et al. (2000a), which (as in the
original implementation of Kylafis) avoids obvious
pitfalls recently highlighted by Lee et al. (2016), while
preserving speed and accuracy, as demonstrated in Popescu \& Tuffs (2013)
and Natale et al. (2014). The 3D code provides an explicit calculation of all
order scattered light. Results have been checked against each other using both 
codes.
The linear resolution of the calculations was up to 25\,pc, which is easily
  sufficient to model the resolved latitude profiles for structures at the
  galactic center. In addition the data, which was highly resolved, showed no
additional structure (e.g. a thinner layer in z) with sizes below the 
resolution of the code. For the
optimisation of the infrared radiation fields the relevant angular
resolution is that of IRAS and Planck bands, which is 5 arcmin,
corresponding to a linear resolution of approx. 12 pc at the galactic
centre. The equivalent
numbers for COBE (tracing direct stellar light from old stars in the NIR/MIR)
is around 40 arcmin/90 pc.
 
The optical constants (from UV to submm) of the dust model used in the
computations were those of Weingartner \& Draine (2001) and Draine \& Li 
(2007), whose grain model incorporated a mixture of silicates, graphites and 
PAH molecules. These optical constants are appropriate to model diffuse 
interstellar dust in the Milky Way, as Draine \& Li (2007) optimised the 
relative 
abundances and grain size distributions of the chemical constituents to fit the 
extinction law and infrared/submm emissivity of translucent high latitude 
Cirrus dust clouds in the solar neighbourhood. The model for the dust 
emission incorporates a full calculation of the stochastic heating of small 
grains and PAH molecules. As described in PT11, our model 
accounts for possible variations in the infrared/submm emissivities of grains 
in dense opaque molecular dust clouds by employing dust emission templates 
empirically calibrated on observed infrared/submm emission spectra when 
accounting for the emission from such structures. 
\label{sec:model}
\begin{figure}
\centering
\includegraphics[scale=0.50]{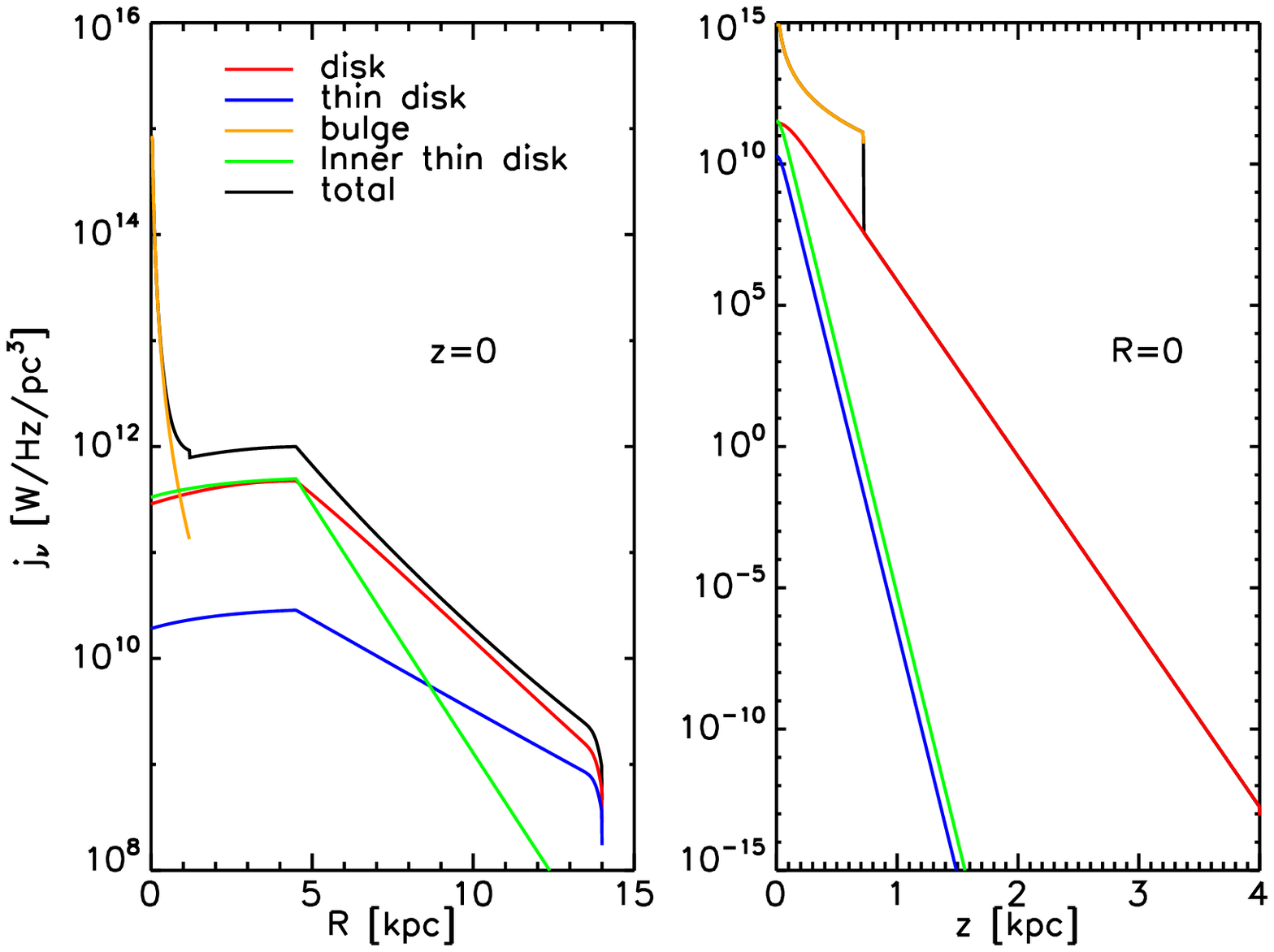}
\includegraphics[scale=0.50]{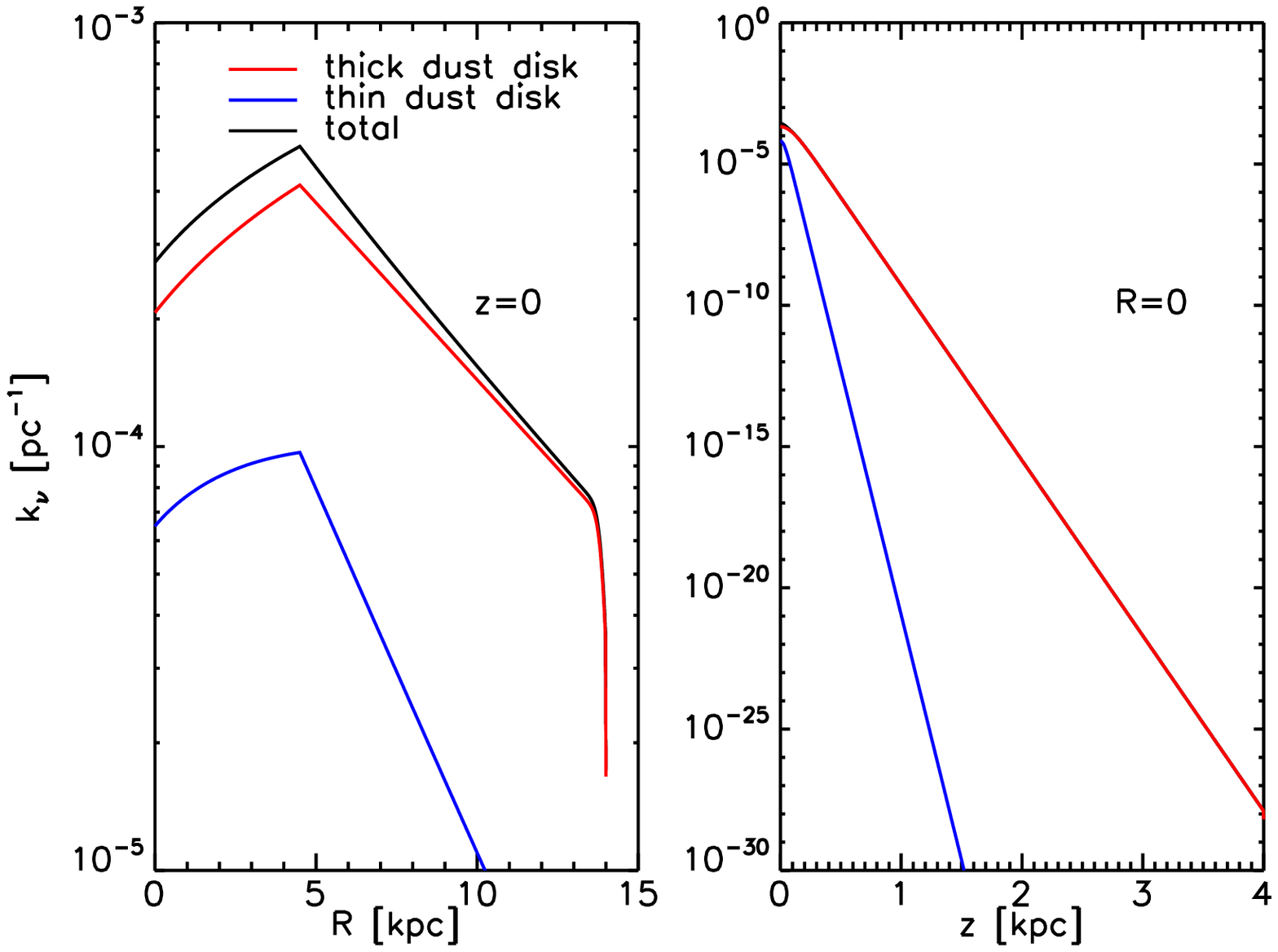}
\caption{Top: Radial (left) and vertical (right) profiles of stellar emissivity
in the K band used in our model. The radial profile is for $z=0$ kpc and the
vertical profile is for $R=0$ kpc. The different colours are for the 
stellar components of the model: disk (red), thin disk (blue), bulge (orange),
inner thin disk (green). Bottom: Same for dust opacity. The red line is for the
thick dust disk and the blue line for the thin dust disk. The total
emissivity/opacity is plotted in all panels with black.}
\label{fig:emissivities} 
\end{figure}

As in our model for external galaxies, the large-scale
distribution of stars consists of an old stellar disk (``the disk''), emitting in the optical/NIR, a UV- and optically/NIR-emitting young stellar disk (``the thin disk''), and  an optically/NIR-emitting bulge.
In addition, to reproduce the observed latitude and longitude profiles in the inner disk in Milky Way, we found that we 
needed to invoke a third component of stellar emissivity, consisting of both a very young 
stellar population, traced by the observed
24\,${\mu}m$ emission in the inner disk, as well as an intermediate-age 
population, as seen in the NIR. We refer to this  component of stellar emissivity as the 
``inner-thin disk''. In our axisymmetric model, the bulge component is
used to account for the combination of both the bulge and the inner bar of the
Milky Way, despite the geometry of the latter. All the stellar components are seen through a 
common distribution of diffuse dust. The diffuse dust is 
distributed into two disks associated with the old and young stellar 
populations, which we refer to as the \lq\lq thick\rq\rq\, and \lq\lq thin dust disk\rq\rq\, 
respectively,  on account of the different thickness of these structures. Thus in our model of the Milky Way the 
stellar volume emissivity and the dust density distributions for all the disk 
components $i$ are described by the following generic formulae: 

\begin{equation}
\label{eq:model}
w_i (R,z|h_i,z_i) = \\
\begin{cases} 
{\displaystyle
A_o\left[\frac{R}{R_{\rm in}}(1-\chi) + \chi\right]\frac{z_i(0)}{z_i(R)}\exp{\left(-\frac{R_{\rm in}}{h_i}\right)}}\\
{\displaystyle 
\times \rm{sech}^2{\left(\frac{z}{z_i(R)}\right)}, 
\hspace{1.8cm} {\rm if}\hspace{0.1cm} R < R_{\rm in}}\\\\
{\displaystyle
A_o\frac{z_i(0)}{z_i(R)}\exp{\left(-\frac{R}{h_i}\right)}}\\
{\displaystyle
\times\rm{sech}^2{\left(\frac{z}{z_i(R)}\right)},
\hspace{1.8cm} {\rm if}\hspace{0.1cm} R \geq R_{\rm in}
}
\end{cases}
\end{equation}

\noindent
with: 
\begin{gather}
\label{eq:flare1}
z_i(R) = z_i(0) + \left(z_i(R_{\rm in})-z_i(0)\right)\left(\frac{R}{R_{\rm
    in}}\right)^\gamma \\ 
\label{eq:flare2}
\gamma = \log\left({\frac{z_i(R_\odot)-z_i(0)}{z_i(R_{\rm in}) - z_i(0)}}\right)/\log\left({\frac{R_\odot}{R_{\rm in}}}\right)
\end{gather}
\noindent
where
$h_i$ is the scale length, $z_i(R)$ is the scale height dependent on the radial
distance $R$, $A_o$ is a constant determining the scaling of $w_\nu(R,z)$, 
$\chi$ is a parameter determining the ratio  $w_\nu(0,z)/w_\nu(R_{\rm in},z)$, 
$R_{\rm in}$ is a inner radius, and $R_\odot$ is the galactocentric distance of the Sun, taken to be 8\,kpc. 
One main difference between the generic formula 
for the Milky Way given in Eq.~\ref{eq:model} and that for external galaxies
is that all stellar and dust components have, inside an inner radius 
 $R_{\rm in}$, a decrease in their emissivities/opacities with 
decreasing radial distance, such that the emissivity/opacity at $R=0\,$kpc is 
a fraction $\chi$ of that at $R_{\rm in}$. 
Outside the inner radius the stellar and dust
emissivity/opacity of the disks are described by exponential functions in radial
direction. As in our modelling of external galaxies, we found a
scalelength of the thick dust disk $h_d$  that is larger than that of the old
stellar disk, $h_s$. This is also in accordance with the fact that external
galaxies were found to have dust associated with extended HI disks, beyond the
edge of the optically emitting disks (Popescu \& Tuffs 2003, Bianchi \&
Xilouris 2011). 
For the vertical distribution we allow for a flare, by 
considering a general expression for $z_i(R)$ as given in
Eq.~\ref{eq:flare1},~\ref{eq:flare2}. 
\noindent
Thus, the generic function $w$ of Eq.~\ref{eq:model} is parameterised as:\\
$w_s^{\rm disk}=w(R,z|h_s^{\rm disk},z_s^{\rm disk})$ - for the stellar disk\\
$w_s^{\rm tdisk}=w(R,z|h_s^{\rm tdisk},z_s^{\rm tdisk})$ - for the thin stellar disk\\
$w_s^{\rm in-tdisk}=w(R,z|h_s^{\rm in-tdisk},z_s^{\rm in-tdisk})$ - for the inner thin stellar disk\\
$w_d^{\rm disk}=w(R,z|h_d^{\rm disk},z_d^{\rm disk})$ - for the thick dust disk\\
$w_d^{\rm tdisk}=w(R,z|h_d^{\rm tdisk},z_d^{\rm tdisk})$ - for the thin dust disk\\

For the stellar bulge we used a Sersic distribution with Sersic index 4, whose 
stellar volume emissivity $j_\nu(R,z)$  is defined as:   

\begin{equation}
\label{eq:bulge}
j_\nu(R,z) = j_\nu(0,0) m^{\frac{-(2n-1)}{2n}}\exp\left(-C m^{1/n}\right) 
\end{equation}
with:
\begin{equation}
m = \frac{\sqrt{R^2 + z^2(a/b)^2}}{R_{\rm e}}
\end{equation} 
and $C= 7.67$ for $n=4$.

Examples of stellar emissivity profiles in the J band and of dust opacity for
the different stellar and dust components of our model for the Milky Way are
shown in Fig.~\ref{fig:emissivities}. 
A summary of the values of the geometrical
parameters of the model that are constrained from data is given in 
Table~\ref{tab:geom} and those that are adopted from PT11 are given in 
Table~\ref{tab:geom1}. The detailed
explanation of how the values of these parameters were derived is given in
Natale et al. (in prep).

In addition to the large-scale distribution of diffuse dust and stars, the geometrical
model also incorporates a clumpy component of dust physically associated with 
highly opaque, small filling factor birth clouds in the star-forming
complexes.  This clumpy component is taken to follow the same 
large scale spatial distribution as the UV-emitting stellar populations in each
of the two geometrically thin UV-emitting disks required to model the Milky Way.
Following PT11, a "clumpiness factor" $F$   is determined 
for each of these two UV-emitting disks. $F$ is the total fraction of the non-ionising light 
which is locally absorbed in the star-forming regions where the stars were born, giving rise to 
a local source of warm dust emission (i.e. typically prominent in the 25 and
60\,${\mu}m$ bands). The thin disk was found to have a very high escape
fraction of UV light from the star-forming
regions, meaning the clumpy component of dust emission is negligible ($F=0$).
On the other hand the inner thin disk was
found to have a low escape fraction of UV light, with $F=0.4$, so this geometrical
component of dust emission is dominated (in luminosity terms) by the clumpy component
of dust. Since the degree of local obscuration of
UV light by the birth clouds can be quantitatively modelled in terms of the age of a UV-emitting stellar
population (Tuffs et al. 2004), this provides new insight on the joint distribution over time and
galactocentric radius of the recent star formation activity in the Milky Way.
This is discussed further in Natale et al. (in prep).

\begin{table}
\caption{The geometrical parameters of the model that are constrained from 
data. All the length parameters are in units of kpc.}
\begin{tabular}{lll}
\hline\hline
&parameter & uncertainty\\
$R_{in}$                          & 4.5 & $5\%$\\
$\chi$                            & 0.5 & $20\%$\\
$h_{\rm s}^{\rm disk}(J,K,L,M)$           & 2.20, 2.20, 2.6, 2.6 & $25\%$\\
$z_{\rm s}^{\rm disk}(0, R_{in}, R_{\odot})$ & 0.14, 0.17, 0.30 & $25\%$\\
$h_{\rm s}^{\rm tdisk}$            & 3.20 & $30\%$\\
$h_{\rm s}^{\rm in-tdisk}$            & 1.00 & $+50\%$-$30\%$\\
$z_{\rm s}^{\rm in-tdisk}(0, R_{\odot})$  & 0.05, 0.09 & $15\%$\\
$h_{\rm d}^{\rm disk}$             & 5.2 & $15\%$\\
$z_{\rm d}^{\rm disk}$             & 0.14 & $15\%$\\
$R_{eff}$                               & 0.4 & $20\%$\\
$b/a$                                  & 0.6 & $20\%$\\
\hline
\end{tabular}
\label{tab:geom}
\end{table}

\begin{table}
\caption{Constraints on geometry, following 
PT11.}
\begin{tabular}{ll}
\hline\hline
$h_{\rm s}^{\rm disk}(B)$/$h_{\rm s}^{\rm tdisk}$ & 1 \\
$h_{\rm s}^{\rm disk}(V,I)$/$h_{\rm s}^{\rm disk}(B)$      & 0.97, 0.80\\
$z_{\rm s}^{\rm tdisk}$/$z_{\rm s}^{\rm in-tdisk}$  & 1 \\
$h_{\rm d}^{\rm tdisk}$/$h_{\rm s}^{\rm tdisk}$  & 1\\
$z_{\rm d}^{\rm tdisk}$/$z_{\rm s}^{\rm tdisk}$ & 1 \\
$R_{\rm trunc}$/$h_{\rm s}^{\rm disk}(B)$ & 4.4\\
\hline
\end{tabular}
\label{tab:geom1}
\end{table}


The values of the clumpiness factors $F$ for the two UV-emitting disks
are two of the global parameters of the Milky Way in
our model. The others are the total face-on dust opacity in 
the B-band at the inner radius $\tau^f_B$($R_{\rm in}$), the opacity ratio 
between the thick and thin dust disks $\tau^{\rm f,disk}_B/\tau^{\rm f,tdisk}$, 
the normalised luminosities of the thin disks, expressed in terms of a 
star-formation rate $SFR$ for each disk (Eqs. 16, 17, 18 from PT11), and the normalised 
luminosity (relative to some standard luminosity defined in PT11 to be 10 times the
luminosity of the non-ionising UV photons produced by a 1 M$_{\odot}$/yr young
stellar population) of the old stellar population ``old'' (Eq. 12 from PT11).
We note that it is the combination of $F$ and $SFR$ for each of the thin disks
that controls the amplitude of the UV radiation fields in the
diffuse ISM arising from each disk. In total our model has the 11 
geometrical parameters listed in Table~\ref{tab:geom} (of which $h_s^{\rm disk}$
was independently determined in the J,K,L and M bands, and $z_{\rm s}^{\rm
  disk}$ and $z_{\rm s}^{\rm in-tdisk}$ are optimised as a function of radial 
position describing the flare or taper) and 8 amplitude parameters 
($\tau^{\rm f,disk}_B$, $\tau^{\rm f,tdisk}$, $SFR^{\rm tdisk}$, 
$SFR^{\rm in-tdisk}$, $F^{\rm tdisk}$, $F^{\rm in-tdisk}$, $old$, $B/D$) 
that were constrained from data.

The best fit to the observed maps of the Milky Way provided a solution with 
$\tau^f_B(R_{\rm in})=1.48$, $\tau^{\rm f,disk}_B/\tau^{\rm f,tdisk} =
5.2$, a total star formation rate from the two thin disks of $1.25
M_{\odot}/yr$, and $old=0.35$. 
We adopt throughout this work the position of
the Sun to be in the midplane at a galactocentric radius of 8.0 kpc. 

\begin{figure*}
\centering
\includegraphics[scale=0.49]{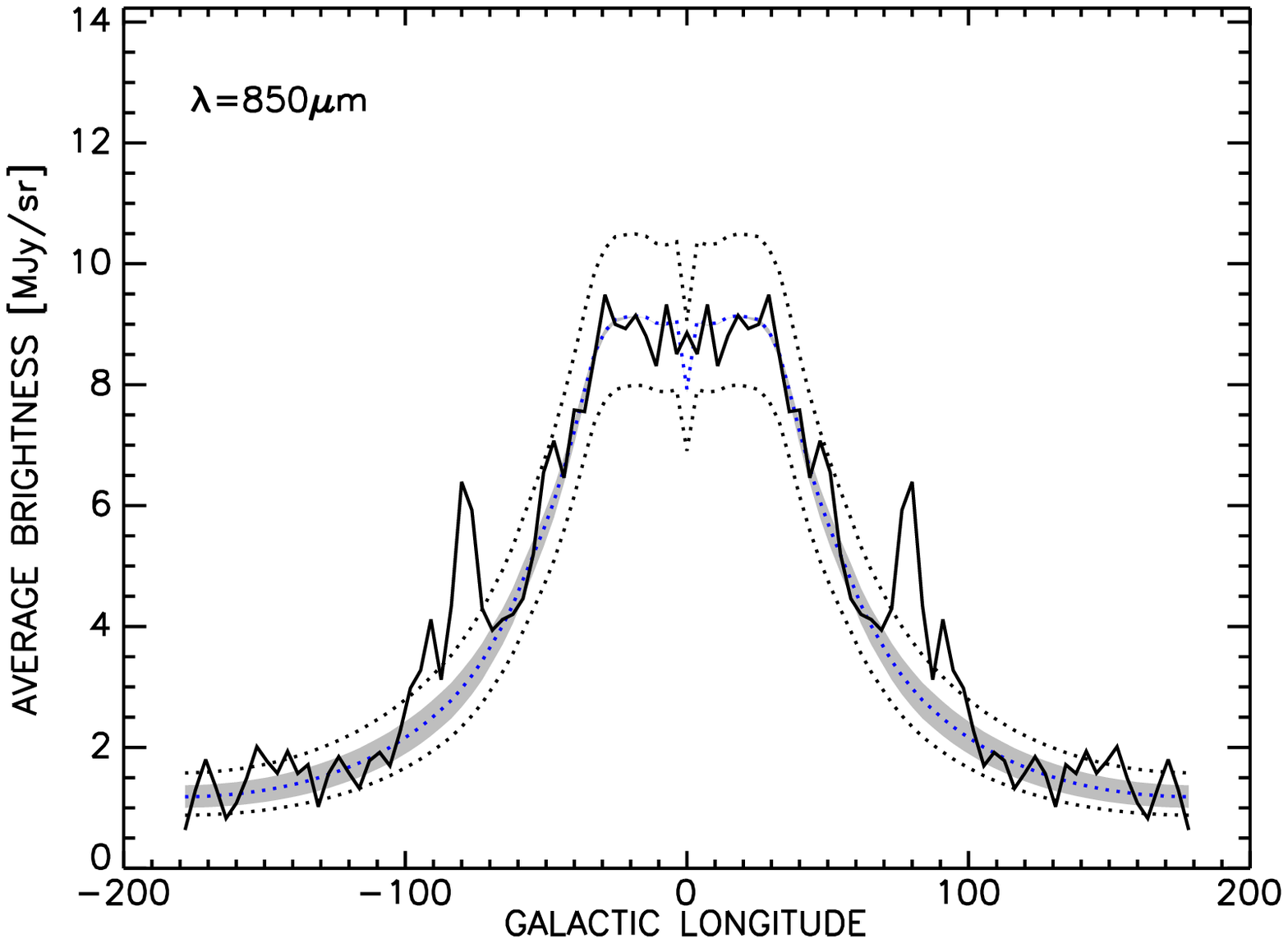}
\includegraphics[scale=0.49]{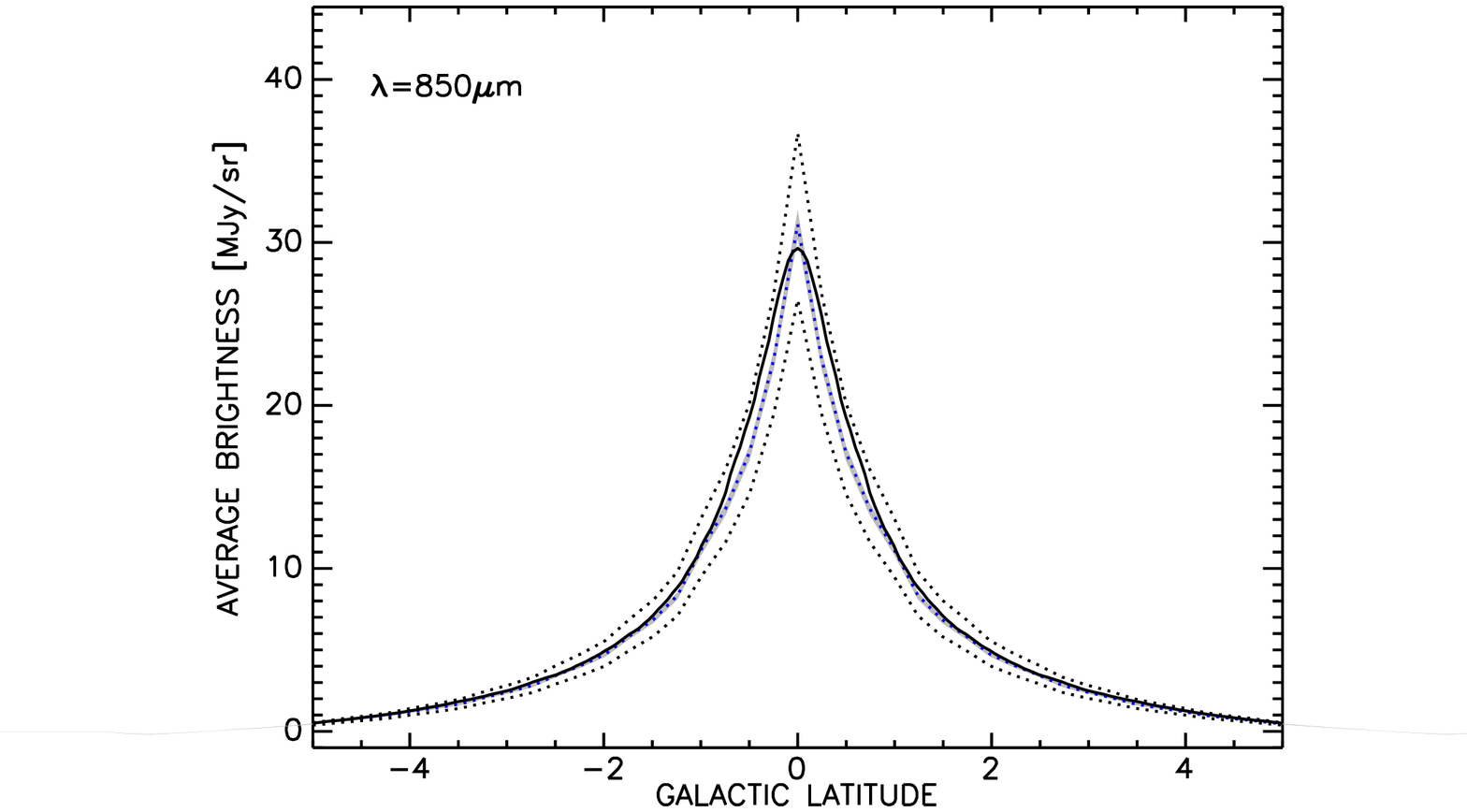}
\caption{Variation in the averaged longitude and latitude model profiles 
(dotted lines) of surface-brightness at 850\,${\mu}$m due to $15\%$ variation 
in the $h_d$. The corresponding observed profiles are plotted with a solid 
line. The shaded area represents the variation
  in the models after the same change in $h_d$ but this time accompanied
  by a change in the $\tau^f(B)$ parameter, such that the centre region of the 
  850\,${\mu}$m longitude profile is fitted. This is equivalent to the
  conditional probability analysis conducted to find errors in $h_d$ and
  $\tau^f(B)$.}
\label{fig:errors_sb}
\end{figure*}
\begin{figure}
\includegraphics[scale=0.5]{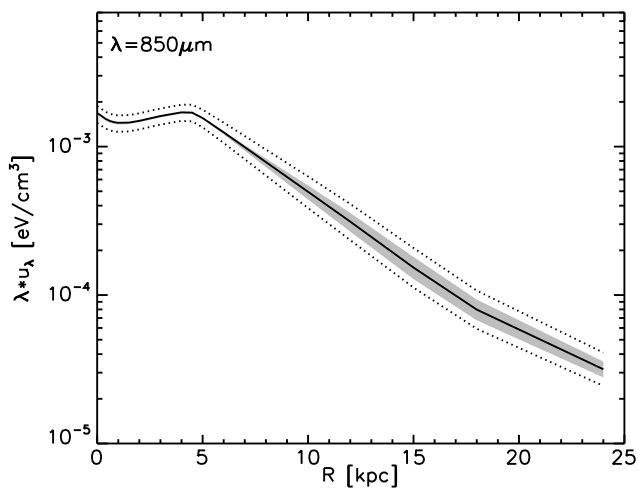}
\caption{Variation in the radial profile of radiation fields (dotted
    lines) at 850\,${\mu}$m due to the $15\%$ variation in the $h_d$ that
can be tolerated by the fits to the longitude and latitude profiles
at  850\,${\mu}$m. The shaded area represents the variation
  in the models after the same change in $h_d$ but this time accompanied
  by a change in the $\tau^f(B)$ parameter, such that the centre region of the 
  850\,${\mu}$m longitude profile is fitted. The shaded area thus represents
  the uncertainty of the radiation fields at this wavelenegth.}
\label{fig:errors_rf}
\end{figure}

The uncertainties in the main geometrical parameters of the model 
(those that are constrained from data) were derived by looking at the departure from the best
fit model of only one parameter at a time, at the wavelength at which the
parameter was optimised. For example, for the scalelength of the thick dust 
disk, $h_d$,
we show in Fig.~\ref{fig:errors_sb} how the fit to the averaged longitude and latitude 
profiles of surface-brightness changes for a $15\%$ change in $h_d$ 
(dotted lines). Because the large
variation in amplitude was compensated for in the optimisation by a 
subsequent variation in the amplitude parameter, $\tau^f(B)$, we also show the 
variation after the profiles were rescaled to fit the central flat part of the 
longitude profiles (shaded area). The shaded area is then taken to represent
the uncertainty in the model fit. The corresponding variations from the best fit
model in the radiation fields is shown in Fig.~\ref{fig:errors_rf}. 

The uncertainties in the values of the geometric parameters estimated in this
way are summarised in Table~\ref{tab:geom}. Overall we estimate that the 
uncertainty in the main geometrical parameters is of the order $15\%-25\%$, 
with the exception of $h_s^{in-tdisk}$, where the errors are large, $+50\%$ and 
$-30\%$ and for $R_{in}$  where the errors are very small, of the 
order of $5\%$. The errors in $h_s^{in-tdisk}$, optimised at 24\,${\mu}$m, are 
large because $h_s^{in-tdisk}$ is smaller than $R_{\rm in}$, reflecting the
fact that this parameter is the expression of a rather abrupt truncation of 
the luminosity in the inner disk. By contrast, the uncertainty in $R_{in}$ is 
small since this determines the longitude at which the profiles flatten, which 
is a prominent feature of the observed profiles at all available wavelengths.

\section{Comparison between models and data}
\label{sec:comp}

As outlined in Sect.~\ref{sec:model} and described in full detail in Natale et al. (in prep), 
the model for the stellar and dust distribution of the Milky Way was derived by 
fitting the detailed surface brightness distribution of stellar and dust emission from the NIR to 
the submm. To compare the model with data we show here in Sect.~\ref{subsec:predictions_maps}
a comparison between the model prediction for the integrated dust
emission from the disk of the Milky Way at the position of the Sun and the corresponding observed SED. 
In Sect.~\ref{sec:lirf} we also show predictions for the local ISRF at optical/UV 
wavelengths, comparing these with the observational constraints. The fits of the model
to the surface brightness distributions are shown in Natale et al. (in prep). 

\subsection{The dust emission SED as seen from the position of the Sun}
\label{subsec:predictions_maps}

\begin{figure}
\centering
\includegraphics[scale=0.50]{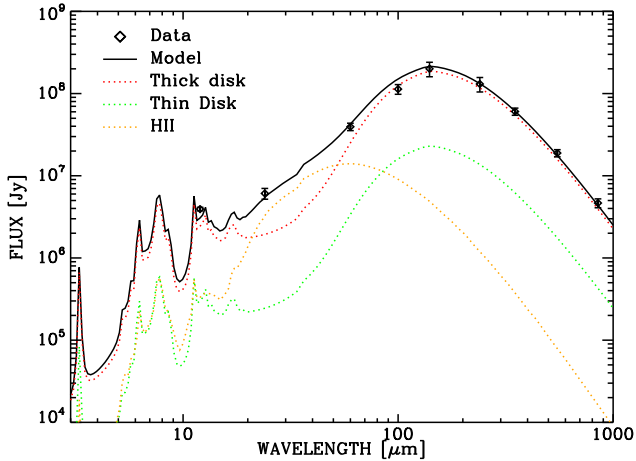}
\caption{Comparison between the radiative transfer (RT) model
predictions (solid line) and 
the infrared data (symbols) for the background-subtracted integrated flux of 
the Milky Way within a 10 degrees strip in galactic latitude ($\pm 5^{\circ}$), as seen from the position of 
the Sun. The different colour dotted-lines represent the different components of
the dust emission model SED: the thick dust disk (red), the thin dust disk
(green) and the clumpy component (orange). The data points at 12,
25, 60 and 100\,${\mu}$m are from IRAS, those at 140 and
240\,${\mu}$m are from COBE, and those at 350, 550 and 
the 850\,$\mu$m are from Planck. The 850\,$\mu$m map has been corrected for
contamination from the CO $J=3\rightarrow 2$ line.}
\label{fig:model_vs_data}
\end{figure}

The infrared volume emissivity distribution derived from the RT 
model of the Galaxy was used to derive the predicted brightness distribution 
maps of the entire sky, as observed from the position of the Sun. Corresponding maps of the observed dust emission were obtained at 12,
25, 60 and 100\,${\mu}$m from IRAS, at 140 and 240\,${\mu}$m from
COBE, and at 350, 550 and the 850\,$\mu$m from Planck. 
In order to avoid contamination with local structure or halo emission, not 
included in the model, both the model and observed maps were then integrated 
over a 10 degree wide strip in galactic latitude (i.e. the same area of sky used for the optimisation)
after subtracting a background from just outside the boundary of the strip in the same manner as 
done from the maps used in the optimisation.
The resulting comparison between the observed and predicted 
background-subtracted dust emission SED of the Milky Way is shown in 
Fig.~\ref{fig:model_vs_data}.
 
The error bars shown in the plot take into account uncertainties in the 
absolute photometric calibration of IRAS, COBE and Planck data, as well as uncertainties 
in the zodiacal light models, but do not take into account possible angular
variations in the background light over the area of the 10 degree strip.
A very good agreement between model and data can be seen, as is to be expected
given that the model gives a good fit to the profiles in galactic latitude and longitude at all
fitted wavelengths.

At 12\,${\mu}m$ our model slightly overpredicts the data, but, as noted in Sect.~2, this band is 
subject to uncertainties in the PAH abundance, and was not included in the optimisation. At 100\,${\mu}m$ the model slightly over predicts the observed flux, but careful examination of the 
detailed surface brightness distribution in the neighbouring bands at 140 and 
60\,${\mu}m$, which are all consistent with the model and between themselves, 
suggests that there may be a small systematic error in the IRAS 100\,${\mu}m$ flux 
calibration where the scan paths of the IRAS survey 
traversed the galactic plane (Natale et al. in prep).

The agreement seen in the submm range justifies the distribution of dust
opacity in the model. The agreement in the MIR, where the dust emission is 
predominantly powered by UV, means that the model prediction 
for the illumination of grains by UV light is also good, both in the diffuse
ISM and locally in the star-formation regions. This is the most direct
constraint of the global distribution of UV light in the Milky Way,
where direct measurement of UV light can only be performed for nearby
stars. Finally, the correct prediction of the colour between the peak of the SED
in the FIR and the MIR shows that both the $SFR$ and the probability of escape
of non-ionizing UV light from star-formation regions in the diffuse ISM,
determined in the model by the $F$ factor, are also correctly
modelled.  While Fig.~\ref{fig:model_vs_data} effectively shows the contribution of the
galactic disk to the local ISRF at the position of the Sun, in Natale et al. (in prep) we  
show that the predictions of our model for the
surface-brightness distribution in the 1.2-850\,${\mu}$m domain also
agrees well with the observed distributions, confirming that the
detailed spatial distribution of dust opacity and stellar emissivity, and thus the ISRF, is
also correctly modelled throughout the volume of the Milky Way.

\subsection{Predictions for the local optical/UV interstellar radiation fields}
\label{sec:lirf}

\begin{figure}
\centering
\includegraphics[scale=0.5]{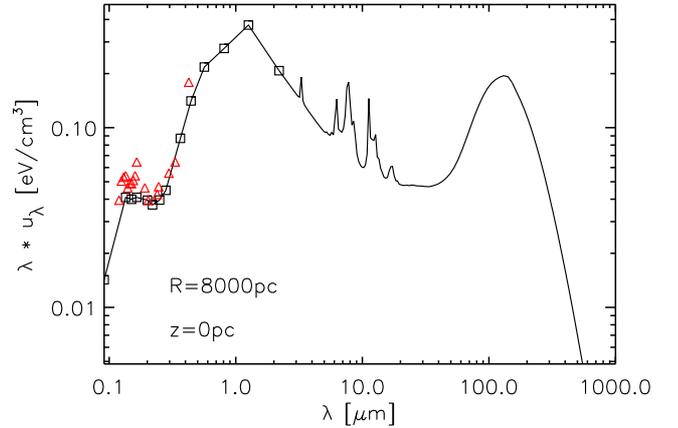}
\caption{Comparison between our radiative transfer model prediction 
(solid line) of the local ISRF from the modelled structures of the Milky Way, 
and the existing observational data. 
The red triangles are the data from Henry et al. (1980) and from Witt et al. 
(1972). The squares are the sampling points of our model for the direct 
stellar light.}
\label{fig:sed_lirf}
\end{figure}

\begin{figure*}
\centering
\includegraphics[scale=0.90]{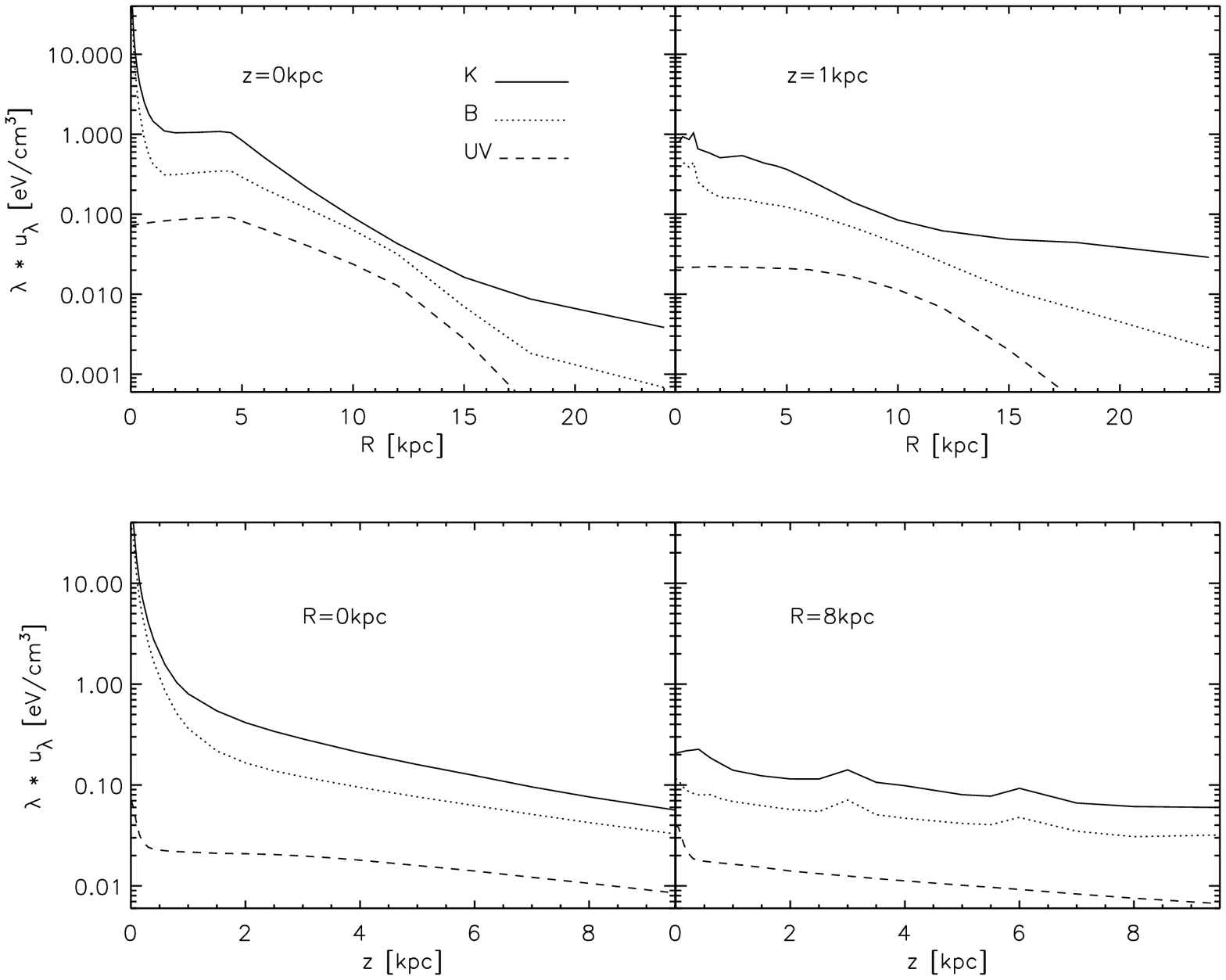} 
\caption{Top: radial profiles of radiation fields in direct stellar light 
at $z=0$\,kpc (left) and
  at $z=1$\,kpc (right). Bottom: vertical profiles of
  radiation fields at $R=0$\,kpc (left) and $R=8\,$kpc. The different lines
  correspond to the profiles at $2.2\,{\mu}$m (solid line), $4430\,{\AA}$
  (dotted line) and $1500\,{\AA}$ (dashed line).}
\label{fig:rf_profiles}
\end{figure*}
\begin{figure*}
\centering
\includegraphics[scale=0.90]{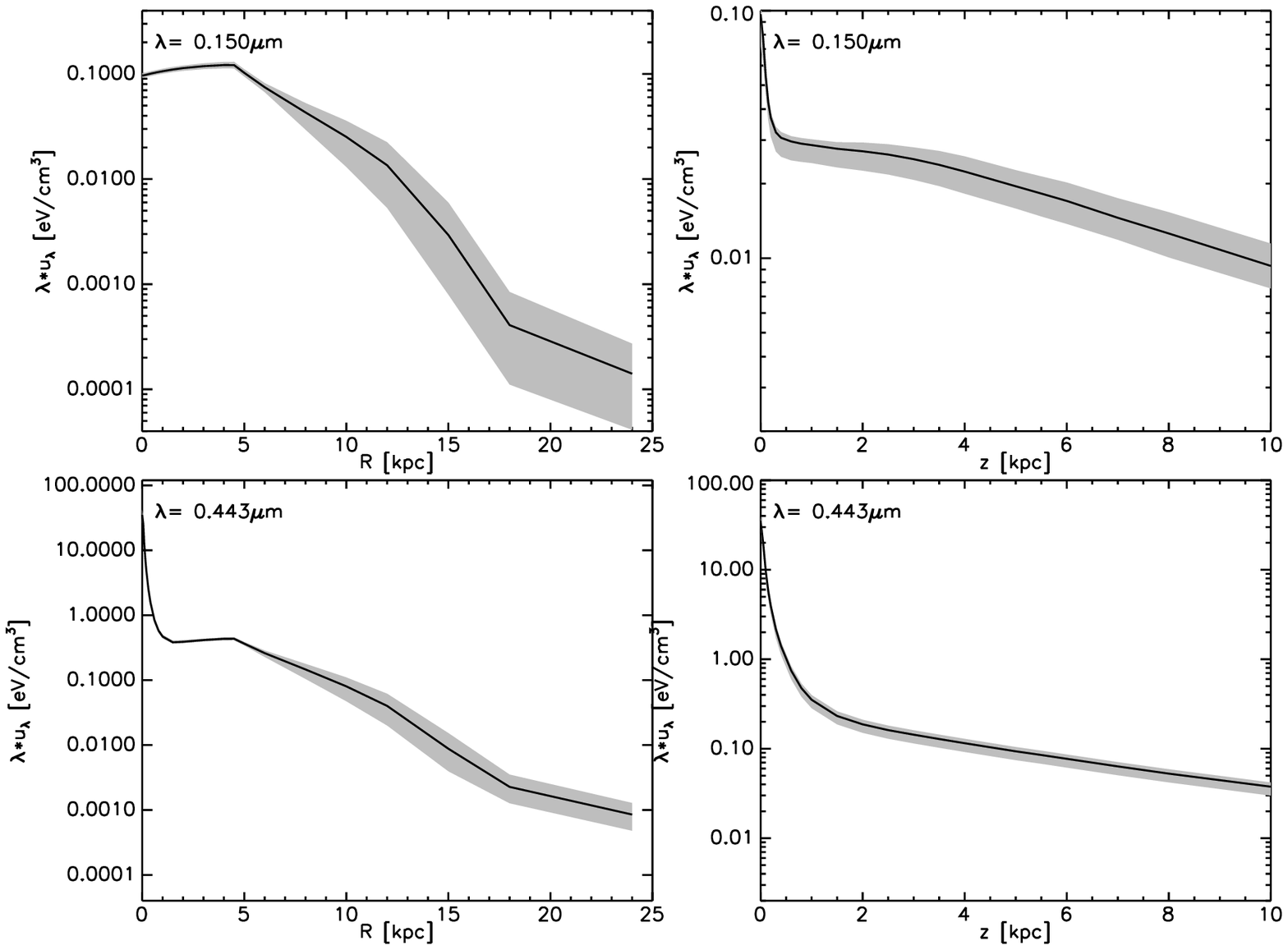} 
\caption{Examples of the uncertainties (shaded area) in the radiation 
fields in direct stellar light arising from the combined uncertainties in the 
geometrical parameters listed in Table~\ref{tab:geom}. The solid line shows 
the radiation fields for UV $1500\AA$ and B band, as plotted in the left hand 
column of Fig.~\ref{fig:rf_profiles}.}
\label{fig:rf_errors}
\end{figure*}

The total intensity and the colour of the local UV/optical ISRF, as parameterized by
Mathis, Metzer \& Panagia (1983; MMP), is considered a benchmark quantity 
in the literature, with many measurements or model calculations
given in units of the MMP ISRF.  Since the 
MMP ISRF is a local measurement, it is very sensitive to nearby structure, in particular for shorter wavelengths where the
range of photons is limited.  It therefore does not provide a strong
constraint of the 
model on global scales, as the infrared maps of the Milky Way do. It is
nevertheless the  only representation of
the UV/optical ISRF based on some direct
observations  and for this
reason we compare the observational data used to constrained the MMP ISRF with 
the predictions of our RT model in
Fig.~\ref{fig:sed_lirf}. This is a relatively challenging test of our determination of
the UV/optical radiation fields, since no UV/optical data was used in our optimisation
analysis. Taking into account the proviso above, there is a good
overall agreement between our model predictions and the MMP local ISRFs.

In passing we note that, in many studies relying on values for the ISRF, the approximation is made
that the shape of the SED of the MMP local ISRF is the same as that of the
SEDs of radiation fields at different galactocentric positions in our Galaxy or in external
galaxies. This is despite the great range of variation in the colour
of the radiation fields with position in a galaxy, as we have shown
in Popescu \& Tuffs (2013) for the PT11 model of external galaxies, and
we now show below for the particular case of the Milky Way.

\section{Results}
\label{sec:results}

\subsection{The radial and vertical profiles of the radiation fields}

For the calculation of the radiation fields we used an irregular grid which is
better suited to sample the different morphological components of the model. In
the radial direction we give the results for 23 positions, with a variable 
spacing ranging from 50\,pc in the inner disk to 6\,kpc at a radial distance of 24\,kpc. In
vertical direction we calculate 22 positions, with a variable spacing
ranging from 50\,pc close to the disk plane and up to 2\,kpc at a distance 
of 10\,kpc above the disk. 

\begin{figure*}
\centering
\includegraphics[scale=0.90]{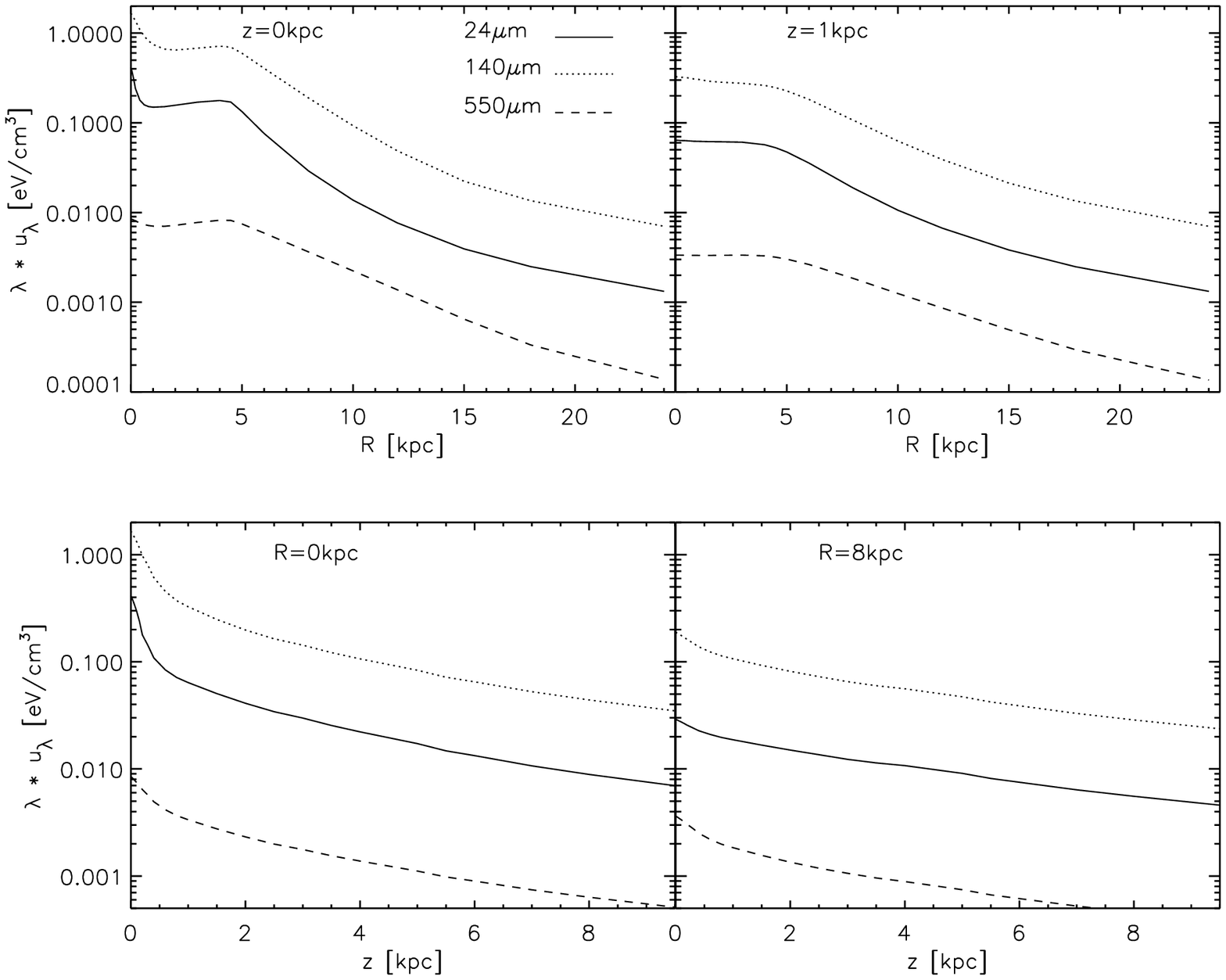} 
\caption{Top: radial profiles of radiation fields in dust emission 
at $z=0$\,kpc (left) and
  at $z=1$\,kpc (right). Bottom: vertical profiles of
  radiation fields at $R=0$\,kpc (left) and $R=8\,$kpc. The different lines
  correspond to the profiles at  $24\,{\mu}$m (solid line), $100\,{\mu}$m
  (dotted line) and $350\,{\mu}$m (dashed line.}
\label{fig:rf_dust_profiles}
\end{figure*}

The radial and vertical profiles of the radiation fields in direct stellar
light in the Milky Way reflect the detailed structure exhibited by the stellar
emissivity of the different stellar components of our model, as well as the 
general characteristics of the fields originating from the three different
vertical structure components emitting stellar photons in our model, the disk, 
the thin disk and the bulge, as described in Popescu \& Tuffs (2013; PT13). 
This can be seen in the
examples given in Fig.~\ref{fig:rf_profiles}, where we plotted the radial 
profile in the plane of the disk and at 1 kpc above the disk, as well as the 
vertical
profile in the centre of the disk and at a radial distance of 8\,kpc. The
profiles were plotted for a NIR, an optical and a UV wavelength.

Inspection of the radial profiles in the plane of the disk (see upper left
panel of Fig.~\ref{fig:rf_profiles}) in the K band shows that the inner 
0.5-5\,kpc is rather flat, due to the decrease in the stellar emissivity and
dust opacity with decreasing radial distance within the inner radius 
$R_{\rm in}$. If the emissivity and opacity profiles were to increase 
exponentially  throughout the centre, then the RFs would be expected to 
decrease monotonically at this wavelength, following the optically thin 
solutions of  the disk described in PT13. In the B-band the
profile within the 0.5-5\,kpc not only flattens, but even decreases with 
decreasing radial distance, due to the additional effect of increasing the
optical depth at this wavelength. As described in Sect. 5.3 of PT13, for a 
pure exponential disk the radiation fields are expected to flatten in the inner
region. The same trend is visible in the UV. In the inner 0.5\,kpc the RF are
dominated by the bulge component.

Outside $R_{in}$ the solutions for the RF follow the general trends predicted
for exponential disks in PT13. In the UV the solutions are dominated by the 
thin disk, with the characteristic shallow monotonic decrease. The break in 
the UV profile at around 14\,kpc is due to the assumed stellar truncation. The 
very prominent break of the UV profile is due to the optically thick character 
of the solution for the thin disk. In the B-band, where the disk is less 
opaque at these radii, the truncation becomes less pronounced in the profiles 
of RFs, with the profile tending to a $R^{-2}$ dependence beyond 18\,kpc. In 
the K-band the solution is optically thin. Therefore, the truncation is not 
visible in the profile, but rather the profile tends smoothly towards the 
$R^{-2}$ dependence.

At 1\,kpc above the plane of the disk (see upper right panel of 
Fig.~\ref{fig:rf_profiles}) the radial profiles of the RFs are only dominated
by the disk and inner thin disk in the K- and B-band, since at that distance 
(more than one $R_{eff}$ of the bulge) the cuspiness in the stellar emissivity 
of the bulge disappears. As expected from PT13, the UV profile becomes rather
flat up to large radii. This is due to the fact that the thick dust disk, 
rather than the thin dust disk, controls the emission escaping from the thin 
disk (see detailed explanation in PT13).  

The vertical profiles in the centre of the Milky Way (see lower left panel of 
Fig.~\ref{fig:rf_profiles}) show again the dominance of the disk and inner thin
disk in the K- and B-bands, except for the inner $\sim 0.5$\,kpc, where the 
cuspiness is due to the bulge. The UV profile shows the general 
characteristics of the thin disk, namely a monotonic decrease above the disk, 
followed by a flat plateau at relatively large vertical distances, which 
eventually steepens to an $R^{-2}$ dependence. The flattening of the profile 
was explained in PT13 as a consequence of the optically thick character of the 
solution and of the two dust disk structure. Thus, a vantage point at larger 
vertical distances above the disk will essentially see a constant surface 
brightness until large radii.

At larger radii (see lower right panel of Fig.~\ref{fig:rf_profiles}) the
vertical profiles are rather flat in all cases, because the typical
distance to the emitting regions, irrespective of whether they are optically
thin or thick, does not change much. The bumps at around 3 and 6 kpc are
  an artifact of the radiative transfer calculations, and could be reduced at
  the expensive of an increase in computation time.

In Fig.~\ref{fig:rf_errors} we show examples of uncertainties in the radiation
fields arising from the combined uncertainties in the geometrical parameters 
listed in
Table~\ref{tab:geom}. In the UV the uncertainties are dominated by the
uncertainty in the parameter $h_s^{\rm tdisk}$, which is optimised at
240\,${\mu}$m, and to a lesser extent by the uncertainty in the parameter 
$h_d^{\rm disk}$, which is optimised at 850\,${\mu}$m. At radii smaller than
$R_{in}$ the uncertainties in the radiation fields are very small, because the
radiation fields are primarily determined by the $\chi$ parameter and the
$SFR$, which are both quite precisely derived. In the B band the
errors are smaller than in the UV, but this is due to the fact that no source of
error was incorporated for the parameter $h_s^{\rm disk}(B)$, since this was
fixed rather than constrained from data.

In the far-infrared and submm the galaxy is optically thin, and therefore the
radiation fields follow similar trends to those seen in the stellar radiation
fields in the K-band. This can be seen in both the radial and 
vertical profiles of dust emission 
(see examples from Fig.~\ref{fig:rf_dust_profiles}). Thus, the radial profiles 
in the plane of the disk outside $R_{in}$ (see top right panel of 
Fig.~\ref{fig:rf_dust_profiles}) show that at all infrared wavelengths the
profiles monotonically decrease with increasing wavelength. 
The remaining profiles in Fig.~\ref{fig:rf_dust_profiles} are again similar
to the corresponding K-band profiles from Fig.~\ref{fig:rf_profiles}. The
vertical profiles from the bottom-right panel of
Fig.~\ref{fig:rf_dust_profiles} show a steep increase as $z$ tends to zero, 
not seen
in the stellar profiles at the K-band wavelength. This is because in an
optically thin case, the smaller the scale-height of the emitters, the
higher the emission closer to the disk plane is. Since most of the dust emission
originates from disks that are thinner as compared with the thicker stellar 
disk in the K-band, it is clear that the radiation fields of dust emission 
will have a higher enhancement close to the plane of the disk.

The uncertainties in the dust emission radiation fields are small in 
relation to those in direct stellar light, similar to those plotted in
Fig.~\ref{fig:errors_rf}. This is because there is a very direct empirical link
between the radiation fields and the IRAS and Planck data, from which the bulk
of the geometrical parameters were optimised. This is also coupled to the fact
that the galaxy is tranparent at the FIR-submm wavelengths.
  
\begin{figure*}
\centering
\includegraphics[scale=0.90]{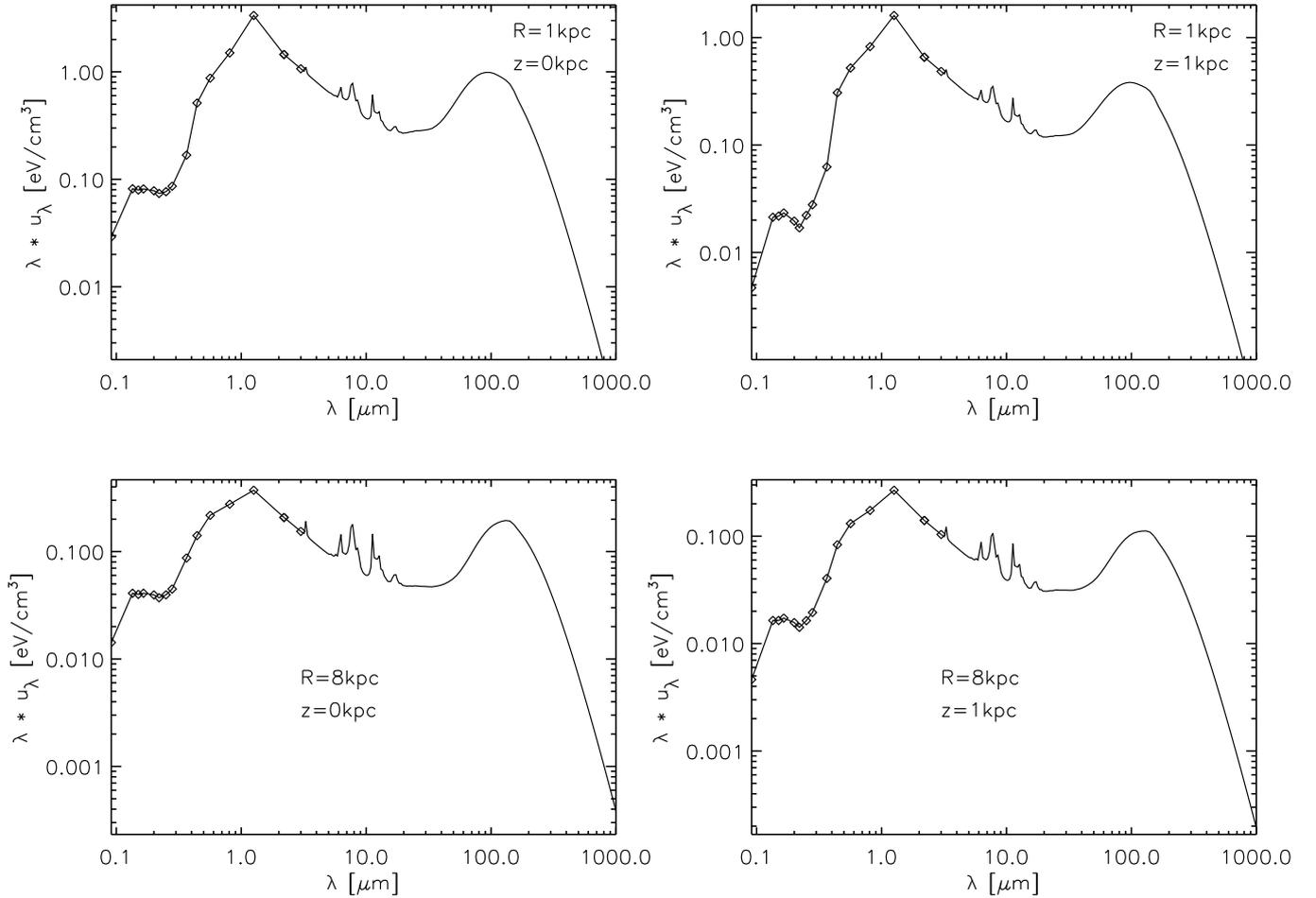} 
\caption{SEDs of radiation fields at various galactocentric positions.The
    symbols are for the UV/optical wavelengths at which the RT calculations were
    performed. For the dust emission the calculations were performed at 500
    wavelengths, logarithmically spaced.}
\label{fig:rf_sed}
\end{figure*}

\subsection{The SEDs of the radiation fields}

One important characteristic of the radiation fields is their spectral
distribution. For applications to high energy
astrophysics, it is particularly important to accurately derive the relative 
proportions of FIR, optical and UV photons, as this
would determine the spectral distribution of gamma-ray photons originating from
Inverse-Compton scattering of the radiation fields. 

To this end we show in
Fig.~\ref{fig:rf_sed} examples of SEDs of radiation fields at different
positions in the Milky Way. It is interesting to note that the SEDs of the RFs
in the centre regions (see top panels) have a strong peak in the near-infrared,
which is due to the prominence of the stellar bulge and of the inner disk, and 
a secondary peak in the FIR. The ratio between the 
NIR and FIR peaks remains approximately constant when going from (1,0)\,kpc to
(1,1)\,kpc above the disk, since both the NIR and FIR radiation fields decrease
by approximately the same factor, as expected due to the optically thin 
character of the solutions in both spectral ranges (see right top panel of both
Fig.~\ref{fig:rf_profiles} and Fig.~\ref{fig:rf_dust_profiles}). 

At larger radial distance in the plane of the disk (see bottom panels of
Fig.~\ref{fig:rf_sed}), the NIR and FIR peaks of the SED become comparable. 
This is due to the fact that the NIR range is no longer affected by the
stellar bulge and the inner disk. For the same reason, the NIR-to-UV ratio is 
also decreased substantially in comparison with the inner parts of the disk.

The FIR SED of the radiation fields peaks at shorter wavelengths in the 
central parts of the disk, due to the strong heating of
the dust in the central regions, including the heating from the bulge and the
inner disk. 

\subsection{Applicability and Limitations of the Predicted Radiation Fields}
\label{subsec:limitations}
The main caveat of our model of the Milky Way is the assumption
   that the properties of the dust grains follow the model of Weingartner 
\&
   Draine (2001). In this model, the absorptivities and emissivities of the
   different grains although empirically anchored through laboratory 
measurements
    in the UV/optical/NIR/FIR range, are quite uncertain in the submm.
  If the submm grain efficiencies were higher than the predictions of the
  WD01
  model, as recently suggested by the Planck Collaboration XXIX (2014),
this
would be equivalent to an overestimation of the UV-optical
optical depth of the galaxy for a model that assumes WD01 dust.
However, in the case of our analysis of the Milky Way, we largely
constrain the amplitude, intrinsic colour, and geometric distribution
of the UV-optical radiation field through the predictions of
amplitude, colour and geometric distribution of the MIR/FIR dust emission.
Thus, the solution for the UV-optical energy
densities themselves will be largely invariant of the choice of dust
model, since, irrespective of whether or not the use of the
WD01 model has led to an overestimation of the
UV-optical optical depth,
one would have exactly the same detailed energy
balance between UV and optical energy absorbed and
infrared energy emitted, and exactly the same spatial structure
in this detailed balance.
This statement holds exactly for the radiation fields in
the UV and optical (B,V,I) range,
since the emission from the Milky Way as a whole at these wavelengths
is anyway inaccessible to direct observation, so has to be
found exclusively through fitting the dust emission data
under the assumption that the geometrical scale ratios
follow the constraints from PT11 (see Natale et al. 2017).
This invariance of the ISRF to the choice of the dust model
will only break down in the NIR, since there we do determine the
amplitude and spectrum of the stellar emission by fitting the NIR
imaging of the directly observed stellar light. However, we also find that 
the UV-optical radiation fields,
as derived from the FIR data, are robust against changes in the parameters
of the NIR emissivity. This is largely because the NIR emission is not 
dominant in the dust heating.
We can therefore conclude that the solutions in the UV/optical
(B,V,I) and MIR/FIR/submm for the radiation fields presented in this paper
are relatively robust against the choice of dust model, but that the NIR
solutions would be altered should the dust model be different. A
dust model with higher submm efficiencies may potentially account for the
underestimation of the J band luminosity in the present solution, but this
would need to be quantitatively investigated in future work. Here we
present the solution for the radiation fields of a Milky Way
having a WD01 dust type.

An obvious limitation of our model is our imposition of axial symmetry, 
which is nonetheless crucial in breaking the luminosity-distance 
degeneracy and in providing a robust way to derive geometrical
distributions independent of prior knowledge coming from stellar counts or gas 
measurements. Clearly we cannot therefore predict variations due to the
boxy-peanut shape of the bulge/bar structure in the inner disk. This may also
account for the underprediction of the J band luminosity, although again this
needs to be quantitatively addressed in future work. Also, we
cannot predict the variations in the diffuse ISRF expected between the arm and
inter arm regions. Tendentially, we 
will under predict the ISRF in the arms, and over predict it in the inter arm regions.
Nevertheless, because the volume of the inter arm regions dominates the volume
of the galaxy,  our solutions should only very slightly over predict the true diffuse ISRF over the
bulk of the volume of the galaxy. 

The boost in the ISRF in the arms will be strongest
in optically thin bands, such as in the NIR/MIR/FIR/submm. In optically thick structures, such as 
spiral arms in the optical, the in situ ISRF will scale as the ratio of emissivity to opacity,
which, in the case of a purely compressive structure like a spiral density wave, should, 
to first order, not affect the in situ diffuse ISRF.  It is also partly because of this that the
emergent SEDs of galaxies modelled with spiral arms is only marginally different to
galaxies with the same global parameters, but modelled as completely smooth disks
(as demonstrated in PT11).

In general, therefore, one can regard 
the model predictions for the ISRF as very close to the true ISRF over most of the volume, whereby the true ISRF will
only strongly exceed the model predictions in spiral arms and the bar in the infrared,
and near powerful discrete sources. In practice, this means that one can use
the model to place lower limits on the inverse-compton emission component of the gamma-ray emission
when modelling individual high energy emission sources. When modelling the global 
inverse compton component of the gamma-ray emission of the Milky Way from scattering off the
ISRF, the model should, analogous to its prediction of the observed longitude and latitude profiles in the infrared, give a systematically correct prediction of the inverse compton emission profiles from smoothly distributed CR electrons in the ISM, subject only to moderate undulations due to
the spiral and bar structure.  We should point out that whereas the model can be used to predict the inverse Compton emission from the general 
population of cosmic rays interacting with the ISRF, it can only be used to place lower limits to the component of IC of any localised sources of cosmic rays 
geometrically associated with SF regions.

\begin{figure*}
\centering
\includegraphics[scale=0.9]{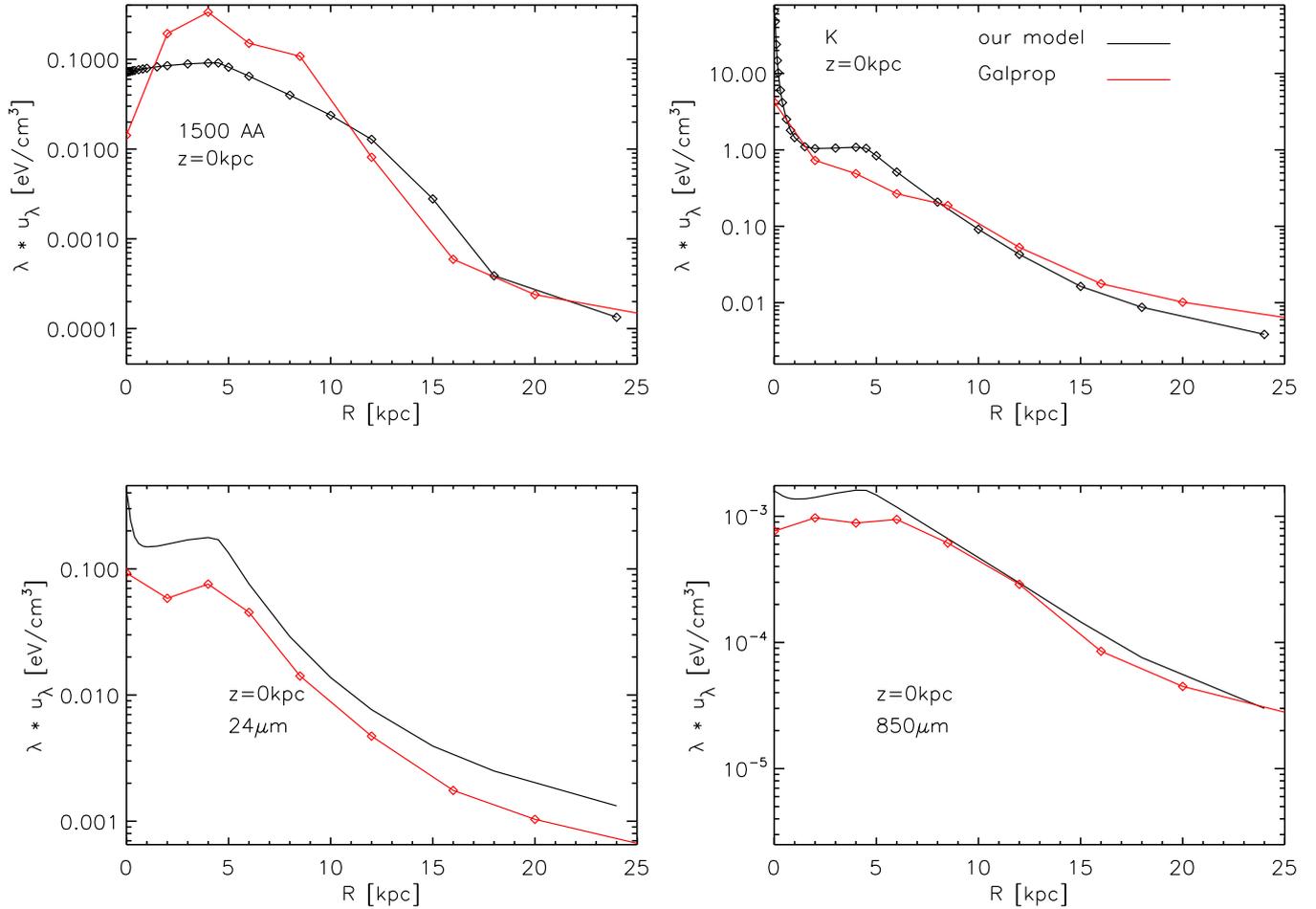}
\caption{Comparison between the mid-plane radial profiles of the
  radiation fields predicted by our RT model (black line) and the GALPROP model
  (red line) for selected wavelengths. The symbols show the sampling
    points in both models}.
\label{fig:comp_galprop_r}
\end{figure*}

\begin{figure*}
\centering
\includegraphics[scale=0.9]{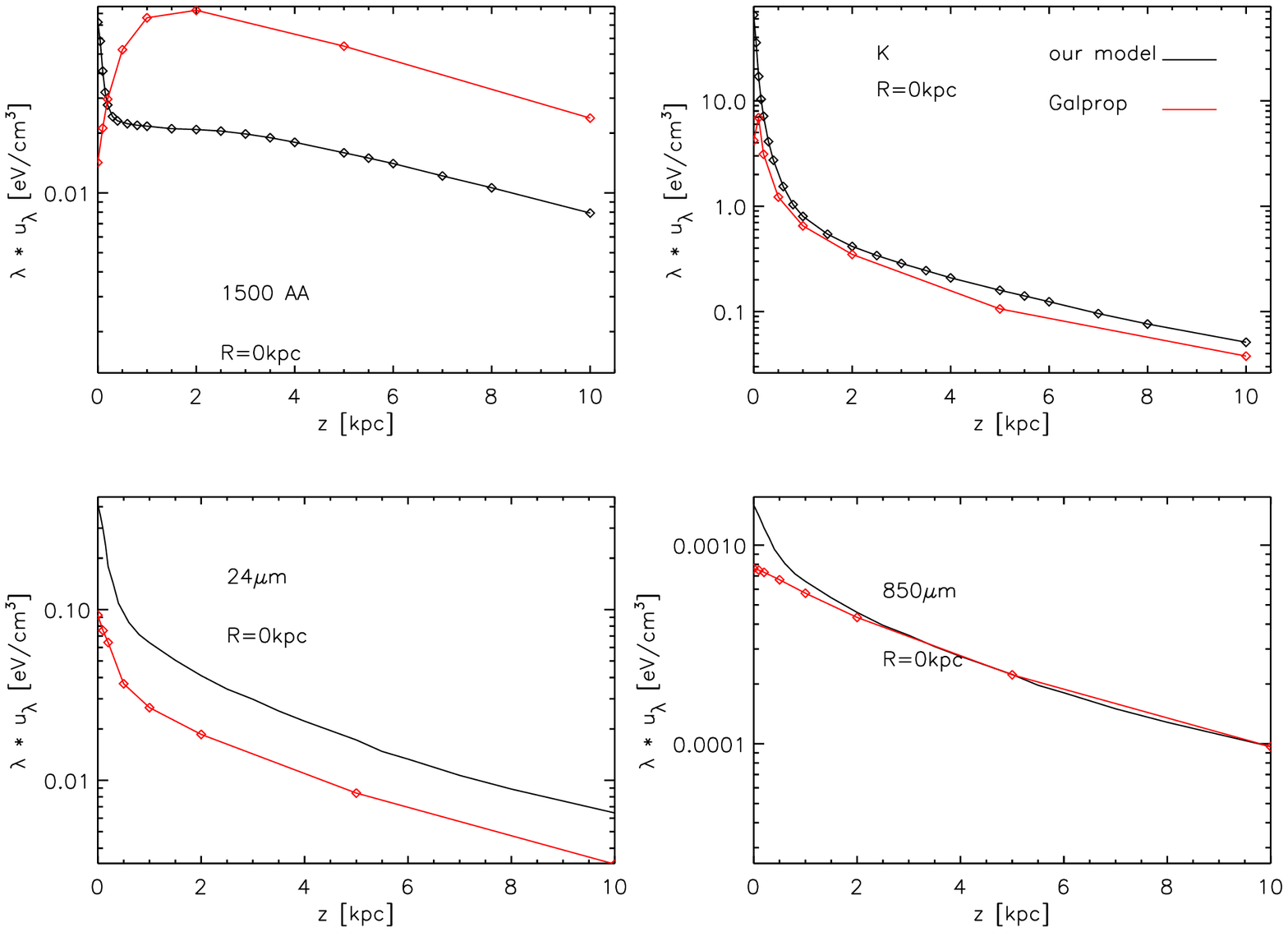}
\caption{Comparison between the central vertical profiles of the
  radiation fields predicted by our RT model (black line) and the GALPROP model
  (red line) for selected wavelengths. The symbols show the sampling
    points in both models}.
\label{fig:comp_galprop_z}
\end{figure*}

Another fundamental assumption of the model is that the dust not 
associated with star-formation regions is diffusely distributed. We know that 
this is actually not the case, since when looking at high resolution Herschel 
images of the plane of the Milky Way one sees that this dust is actually distributed in an intricate pattern of filamentary structure. However, individually these filaments are at a good approximation optically thin to the ISRF, so for the purpose of calculation of transport of photons, one does not need to 
explicitly treat these structures  as clumps. The one caveat to this is that the detailed structure we see at high latitude on the Herschel images are very 
local to the Sun, so do not give direct observational contraints on the 
presence and optical depth of passive small scale structures in the centre of 
the Milky Way. If these structures were optically thick, our model would over estimate the attenuation of starlight from the inner disk.

We also draw attention again here to the fact that our model for the ISRF does not include any component
of ISRF from a possible large scale emission component from a galactic halo, 
where diffuse dust could be collisionally heated with ambient hot plasma (Popescu et al. 2000b).
We refrained from doing this in this work
due to the difficulty in distinguishing between observed smooth emission components extending over the
sky due to local emission above the Sun, and a truly large scale (and thus much
more luminous) halo emission component. The implications of this for the predicted ISRF is in any case negligible in the FIR, because at these wavelengths only a very small fraction of the observed total integrated flux over the whole sky does not arise from the modelled disk structure. However, at 25 micron and shorter wavelengths, a significant (though still minority) fraction of the observed all-sky flux cannot be accounted for by the model for the disk and bulge 
structures of the Milky Way, indicating another origin. Nevertheless, even if originating from a
large halo, this addition MIR luminosity source would significantly boost the
ISRF in the plane of the galaxy only at the solar circle and beyond. The main
observational consequence to gamma-ray astronomy of the existence of a
MIR-emitting galactic halo, should this be shown to exist, would instead be
some increase in the predictions for the pair-production opacity of very high
energy gamma-rays, beyond the predictions we give in Fig.~14 of this paper. We
will return to the nature of the MIR background emission in a future work.

Table~3 (included in the electronic submission only) gives the
values for the total local ISRF at the Solar position, $u_{\lambda, {\rm total}}$,
in each of the bands of COBE/DIRBE, IRAS and Planck, obtained from the
total all-sky flux $S_{{\rm allsky}}$ observed on the maps, calculated
through the relation:
\begin{eqnarray}
u_{\lambda, {\rm total}} = \frac{1}{\lambda^2} \times S_{{\rm allsky}}
\end{eqnarray}

Table~3 also gives the corresponding values for $u_\lambda$ for
the model, $u_{\lambda, {\rm model}}$. The difference
\newline $u_{\lambda, {\rm background}} = u_{\lambda, {\rm total}} -
u_{\lambda, {\rm model}}$ then gives the contribution to the local ISRF
from the background emission which was excluded in our analysis
of background-subtracted strip maps of width $\pm 5$ degrees in latitude.
We will return to the question of whether this background component
is from the galactic halo, rather than being local,
in a future work.

\section{Comparison with the radiation fields predicted
  by the GALPROP model}
\label{sec:comp_galprop}

To date, the GALPROP model has been the only one providing calculations for the
radiation fields in the Milky Way (Strong et al. 2000, Moskalenko et al. 2002,
Porter\& Strong 2006, Moskalenko et al. 2006, Porter et al. 2008). This model 
has been  extensively used by the high energy community for the interpretation 
of the
gamma-ray emission. For this reason we compare the solutions for the ISRF of 
our RT model with the radiation fields from GALPROP.\footnote{The radiation 
fields from GALPROP were 
taken from 
http://galprop.stanford.edu/, using data files from GALPROP v.54. The 
files with the radiation fields are taken from the files 
/ISRF/Standard/Standard\_x\_0\_z\_flux.dat.}
Here we show a few comparative plots for four
selected wavelengths, at 1500 and 22000\,$\AA$ in the UV and NIR, and at 
24 and 850\,${\mu}$m in the mid-IR and submm. In  
Fig.~\ref{fig:comp_galprop_r} we
plot mid-plane radial profiles and in Fig.~\ref{fig:comp_galprop_z} we plot 
central vertical profiles. 
We find that overall there are systematic differences between the GALPROP 
radiation fields and those predicted by our model, in particular in the UV and 
in the mid-IR. Since GALPROP uses the same model for the size 
distribution and optical properties of the grains 
(Weingartner \& Draine 2001; Draine \& Li 2007) as our model, the differences
must be found in the different geometry considered for the stellar and dust
distributions. Indeed, inspecting the recent comparison between the GALPROP
predictions and the IRAS/COBE/PLANCK maps shown in Porter (2016) it appears that the
geometry used in the GALPROP model is not consistent with the observational
data. This reflects the different concepts involved in the two models. Thus, 
GALPROP assumes the stellar and dust distributions to be known and an input to
the model, while our model derives the stellar and dust distributions as an
output of the model. In doing this GALPROP imports distributions derived
from stellar counts and gas measurements, while we optimise the distributions
on the IRAS/COBE/PLANCK maps.

There may be also some technical differences in the modelling itself. Thus, a 
major reason for the difference in the predictions for the MIR emission could
be the treatment of dust 
locally heated by radiation from young stars in the parent molecular clouds of 
star-forming regions. This process, which takes place on parsec scales, is the 
main  contributor to the energy emitted by a galaxy in the 24-70\,${\mu}$m 
(see Popescu et al. 2000a). As outlined in
Sect.~\ref{sec:model}, our RT model 
incorporates this process by calculating the absorption of radiation in an 
ensemble of opaque, but fragmented star-forming clouds.  
The GALPROP model does not take 
into account this process, but only assumes a 
diffuse large-scale distribution of stellar emissivity and dust 
(Porter et al. 2008). Neglecting the mechanism of locally heated dust emission
is thus one potential reason why the bulk of MIR emission from the MW would be
missed by GALPROP.

Another possible reason for the discrepancy in the MIR could come from
differences in the way the diffuse component of the MIR emission is calculated
by the two models. This emission  arises from stochastically
heated small dust grains and PAH molecules. The transition between
emission arising from grains heated to an equilibrium dust temperature and
those heated impulsively depends both on the size of the emitting
particle but also on the strength and colour of the ISRF. This transition 
needs to be therefore self-consistently
calculated for each grain at each position in the galaxy. Our RT model
incorporates a self-consistent calculation of this transition (see
Popescu et al. 2011). However, the
GALPROP model considers a fixed size where this transition occurs 
(Porter et al. 2008),
independent on the strength and colour of the radiation fields. This
could lead to a systematic underprediction of the diffuse MIR emission in the
outer regions of the Milky Way, and an overprediction of the MIR
emission in the inner regions. This could add to the
discrepancies seen in the comparative plots, although this cannot be a major
factor.

Recently Porter (2016) has advocated that the reason the 2D GALPROP model
fails to reproduce the observed maps in the NIR, FIR and submm is due to
the axisymmetric nature of the model, and a 3D model is
proposed instead. While a detailed 3D model is indeed desirable to account for
the detailed inner bulge-bar regions of the Milky Way, our analysis shows that 
an axisymmetric model is well able to reproduce the overall observed 
multiwavelength  large scale surface brightness distributions.

\section{The effect of the radiation fields on the Inverse Compton 
emission and the gamma-ray opacity}
\label{sec:HE}

\begin{figure*}
\centering
\includegraphics[scale=0.90]{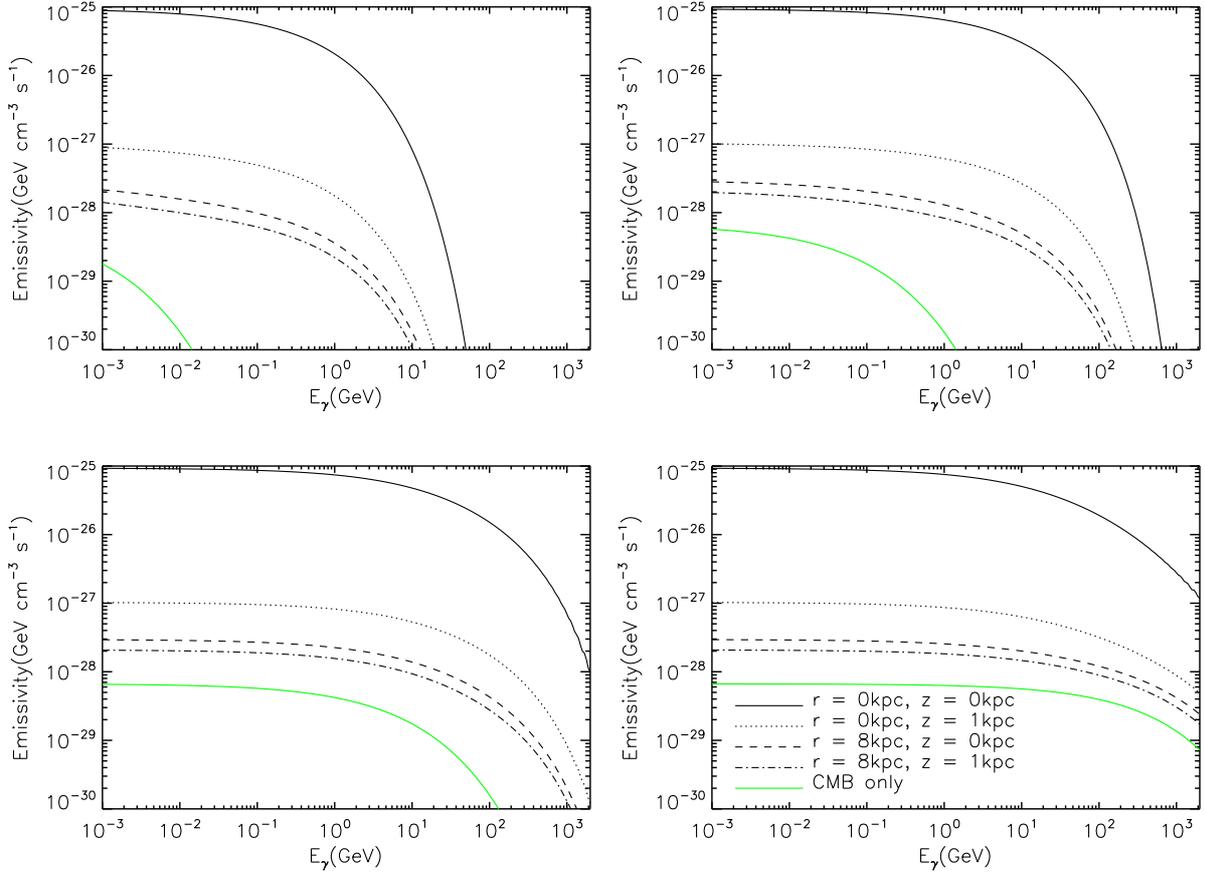}
\caption{Gamma-ray IC emission spectrum for standard electron sources scattered
  by the diffuse ISRF in the Milky Way plus the
  photons from the CMB, at each
  of the four reference positions of Fig.~\ref{fig:rf_profiles}. The electrons 
have a power-law
  distribution $n(E)\sim E^{-3}$, with an exponential cut-off at 
$E_b=0.01$\,TeV (top left), $E_b=0.1$\,TeV (top right), $E_b=1$\,TeV 
(bottom left), $E_b=10$\,TeV (bottom right). As a common reference to each 
panel we also plot the IC spectrum resulting just from the CMB photon field.}
\label{fig:sed_ic}
\end{figure*}

\begin{figure*}
\centering
\includegraphics[scale=0.9]{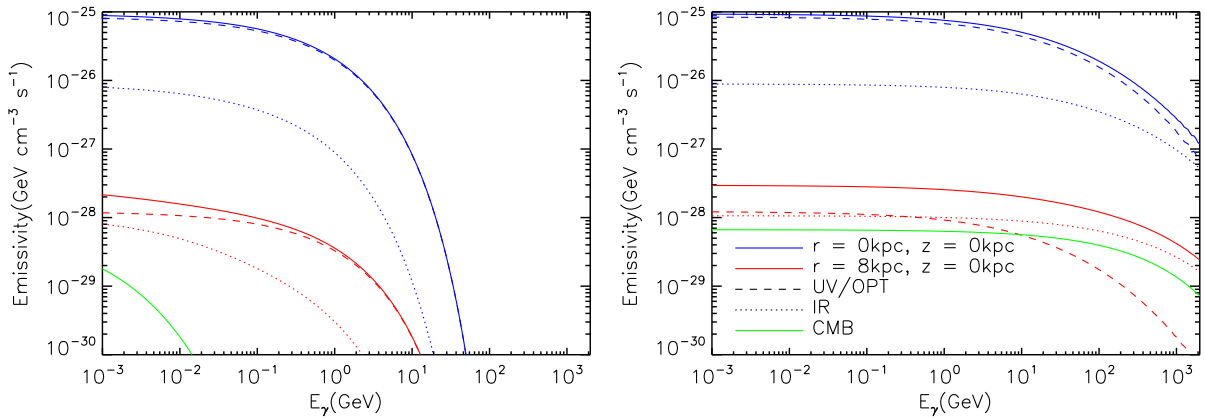}
\caption{Same as in Fig.~\ref{fig:sed_ic}, but also showing the components to 
the IC SED due to infrared photons (dotted line) and UV/optical photons (dashed
line) from the ISRF. The plots are shown only for the
two reference positions in the plane of the disk and only for two cutt-off
energues,$E_b=0.01$\,TeV (left) and $E_b=10$\,TeV (right).}
\label{fig:sed_ic_decomposition}
\end{figure*}

In this section we explore how the spatial and spectral distribution of the 
ISRF in the Milky Way affects the spatial and 
spectral distribution of the gamma-rays produced via Inverse Compton 
scattering, as well as the attenuation of the gamma-rays due to interactions of
the gamma-ray photons with photons of the 
ISRF.  We make
no attempt to model the spatial and spectral distribution of the cosmic ray
electrons, but rather take a standard reference distribution of electron 
energy to illustrate the imprint of the ISRF on the processes
considered and we make the assumption that the electron spectrum is the same
everywhere in the galaxy. In reality the electron spectrum can vary with
position, especially at TeV energies. Moreover, the electrons with energies 
above 10\,TeV cannot propagate more than 100\,pc, therefore their spectrum
should vary with position. However, because in this paper we only want to
illustrate the impact of the ISRF on the predictions for the 
gamma-ray emission, we consider the simplest assumption of a homogeneous 
distribution of electrons. Specifically, we assume the electron spectra at each 
position in the Milky Way to be given by a power law with an index of -3 and 
a high energy exponential cut-off 
$E_b$ ($5.4\times 10^{-12}\,(E/1{\rm GeV})^{-3}\,
\exp(-E/E_{\rm b})\,{\rm cm}^{-2}\,{\rm s}^{-1}\,{\rm GeV}^{-1}$) with a
normalisation  given by the electron flux measured by PAMELA (Adriani et
al. 2011) at 10\,GeV. 

\subsection{Inverse Compton scattering}
\label{subsec:IC}

To calculate Inverse Compton scattering we adopt the formalism described in 
Khangulyan et al. (2014). In the current calculation we neglect the anisotropy 
in the target photon field, although we note that  
anisotropy may cause significant enhancement of the high latitude diffuse
emission (Moskalenko \& Strong 2000). This issue
will be addressed in our future work. 

In Fig.~\ref{fig:sed_ic} we show the predicted gamma-ray IC SEDs for different
positions in the Milky Way, and for different cut-off energies $E_b$ of the 
electron spectra. In each case we also show the predicted gamma-ray IC SED 
resulting just from the Cosmic Microwave Background (CMB) photon field. 

All the IC SEDs are characterised by a break energy $E_{\gamma, b}$, which is
the energy at which a cosmic ray electron with the cut-off energy $E_b$ would 
be scattered on the photons from the radiation fields at the peak of the SED,
having energy $E^{peak}_{ph}$. 
This break energy appears as a knee in the IC SED and is proportional to 
$E_b^2\times E^{peak}_{ph}$.Thus, for the same radiation fields, the break
energy will have lower values for low values of $E_b$ and will increase with
increasing $E_b$. This can be seen in  Fig.~\ref{fig:sed_ic}, where the lowest
values of $E_{\gamma, b}$ are for $E_b=0.01$\,TeV (top left panel), and the
highest values of $E_{\gamma, b}$ are for  $E_b=10$\,TeV (bottom right
panel). One can also see that, for a given cut-off energy $E_b$,the break 
energy of the IC SED powered by the CMB is systematically lower than that of
the IC SED powered by the ISRF, as expected due to the ISRF having a peak at higher energies than the CMB.

For energies beyond the break ($E_{\gamma} > E_{\gamma, b}$) the IC SED is
determined by the most energetic photons in the photon field, with the
steepness of the SED depending on the color of the radiation fields. Thus, the
fall-off  beyond the  $E_{\gamma, b}$ is steeper for the IC SED powered by the
CMB than that powered by the ISRF, since, on the Wien side of the peak, the 
ISRF  have a flatter spectrum than the CMB. One can
also see from Fig.~\ref{fig:sed_ic} that the 
flatter fall-off happens for the IC SED powered by the ISRF at (8,0)\,kpc, since there
the radiation fields have the flattest SED (see Fig.~\ref{fig:rf_sed}).

For energies smaller than the break energy ($E_{\gamma} < E_{\gamma, b}$), the
IC SED is mainly determined by the cosmic ray electron spectrum. Thus, for a
given cut-off energy of the electrons $E_b$, the IC SEDs have a similar shape,
irrespective of the radiation fields powering the gamma-ray emission. This
can be better seen in the bottom right panel of Fig.~\ref{fig:sed_ic}, where
all the SEDs are below the break energy $E_{\gamma, b}$ and therefore are 
parallel to each others. Another feature of the emission is that the
IC flux from the ISRF seed fields always exceeds that
from the CMB fields,
even at large galactocentric radii where the energy density
of the CMB exceeds that of the ISRF. This is because lower energy electrons
are needed to produce a gamma-ray of
some fixed energy by IC scattering from the ISRF, and there are more such
electrons available due to the prescribed power law distribution in energy.

\begin{figure*}
\centering
\includegraphics[scale=0.55]{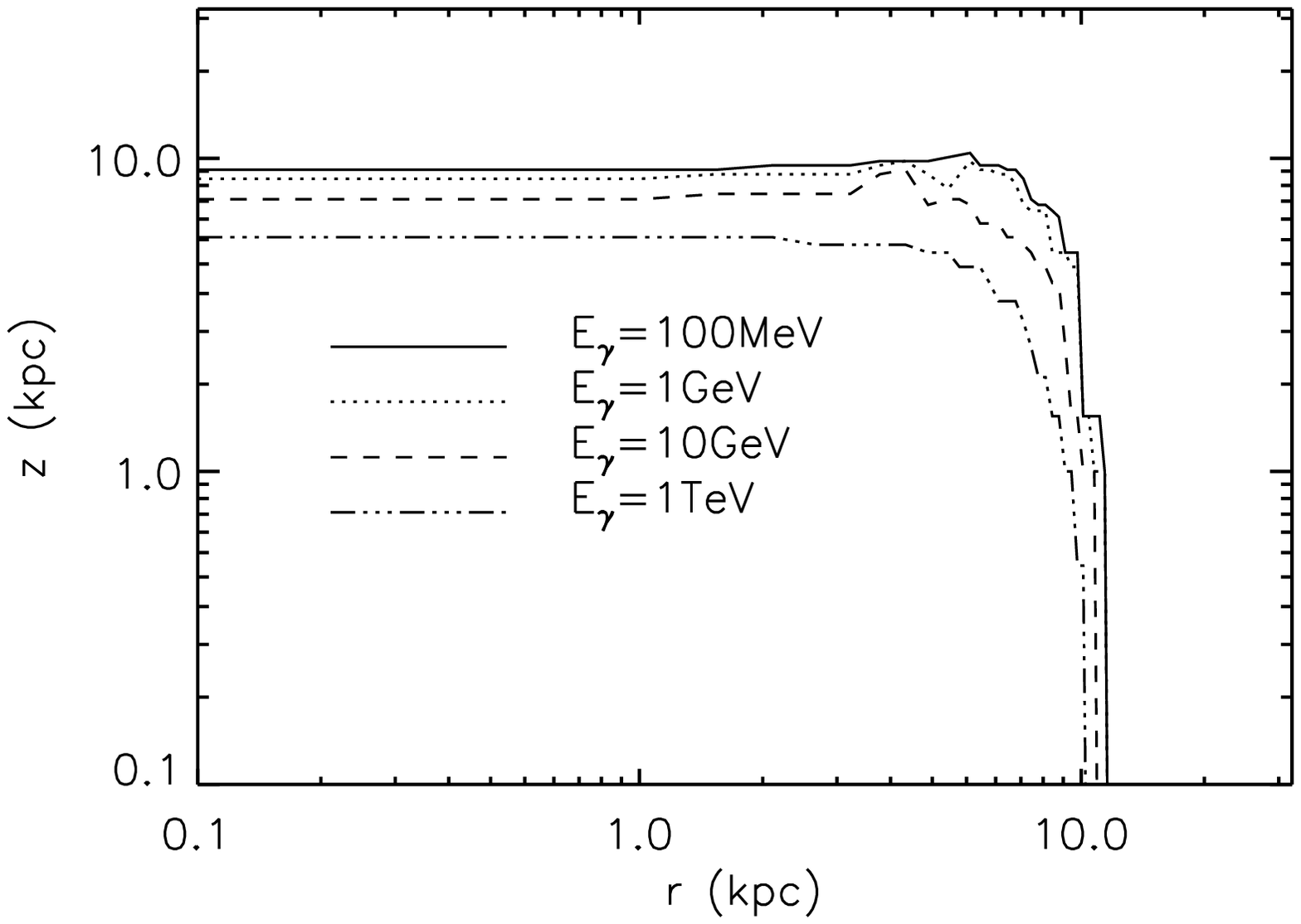}
\caption{Contour plot showing the locus of positions at which the contribution
  from the CMB seed photons to the gamma-ray emission at a particular gamma-ray
  energy is equal to the
  contribution from diffuse ISRF photons. Countours are given in logarithmic 
 steps of 10 in energy for 100\,MeV, 1\,GeV, 10\,GeV, 100\,GeV, 1\,TeV, and 10\,TeV.}
\label{fig:contour}
\end{figure*}

In Fig.~\ref{fig:sed_ic_decomposition} 
we show the IC SED, separated into components arising from the
UV/optical and infrared photons in the ISRF. Thus, for any position in the
galaxy, one can see that for the lower energy
cut-off of the electron spectrum, $E_b=0.01$\,TeV (left panel
of Fig.~\ref{fig:sed_ic_decomposition}), the break energy of the IC SED 
powered by
the UV/optical radiation fields occurs at higher energies than that of the
IC SED powered by infrared radiation fields, which, in turn, occurs at
higher energies than that of the IC SED powered by the CMB. At the higher 
energy cut-off of the electron spectrum, $E_b=10$\,TeV (right panel of
Fig.~\ref{fig:sed_ic_decomposition}), there is a cross-over at high 
$E_{\gamma}$ between the IC SED powered by UV/optical and infrared radiation
fields. This is because the cross-section for the IC scattering of UV/optical
photons enters the Klein-Nishina relativistic regime before the infrared photons.

Another interesting feature of these plots is that for lower energy cut-off of
the electron spectrum, the IC SEDs in the central
regions of the disk have a break at larger gamma-ray energies than that of the
IC SEDs from the outer disk. This is because the IC SED in the central regions is
mainly powered by UV/optical ISRFs (see left panel of
Fig.~\ref{fig:sed_ic_decomposition}), while the IC SED in the
outer regions has approximately equal contributions from the UV/optical and 
infrared radiation fields. 

Overall,  Figs.~\ref{fig:sed_ic} and \ref{fig:sed_ic_decomposition} show that
the IC SED powered by the ISRF dominates over the IC SED powered by the CMB,
for all plotted positions. This dominance prevails within a $\sim 10$\,kpc
radius sphere around the Milky Way, as seen from the contour plots from
Fig.~\ref{fig:contour}. It should be noted that the radius of the sphere
is expected to depend on the hardness of the CR electron energy spectrum.
For CR electrons with a steeper spectrum than the fiducial power-law
distribution $n(E)\sim E^{-3}$ that we have assumed, the ISRF component of
IC emission will continue to dominate beyond $10$\,kpc, whereas for
CR electrons with a flatter spectrum, the cross-over point will
be interior to $10$\,kpc.

\begin{figure*}
\centering
\includegraphics[scale=0.9]{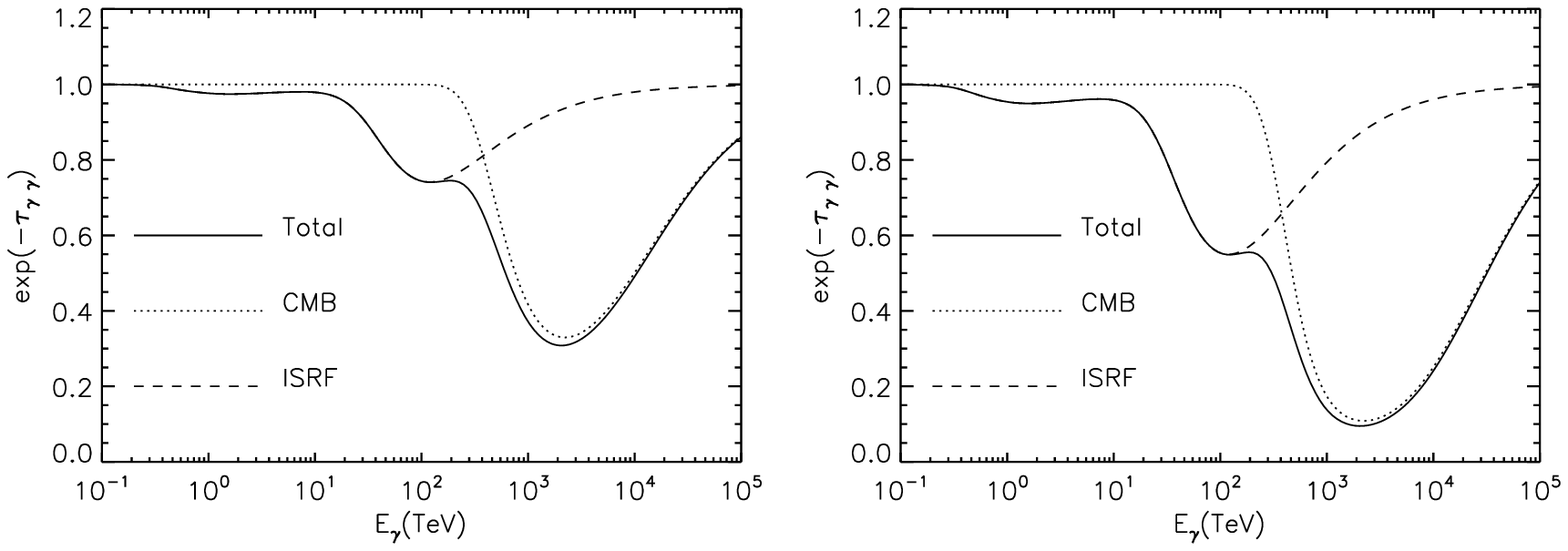}
\caption{The transmission curve (i.e. the fraction of 
photons reaching the Sun as a function of photon energy) for gamma-ray sources 
located at the Galactic Centre (left), and in the plane of the galaxy at a 
location 8\,kpc from the Galactic Centre in the opposite direction to the Sun
(right). The Sun is assumed to be located at $R_{\odot} = 8$\,kpc and 
$z_{\odot} = 0$\,kpc. Both the contribution from the CMB and the ISRF are shown 
in the figure.}
\label{fig:tau_galactic}
\end{figure*}


\begin{figure*}
\centering
\includegraphics[scale=0.9]{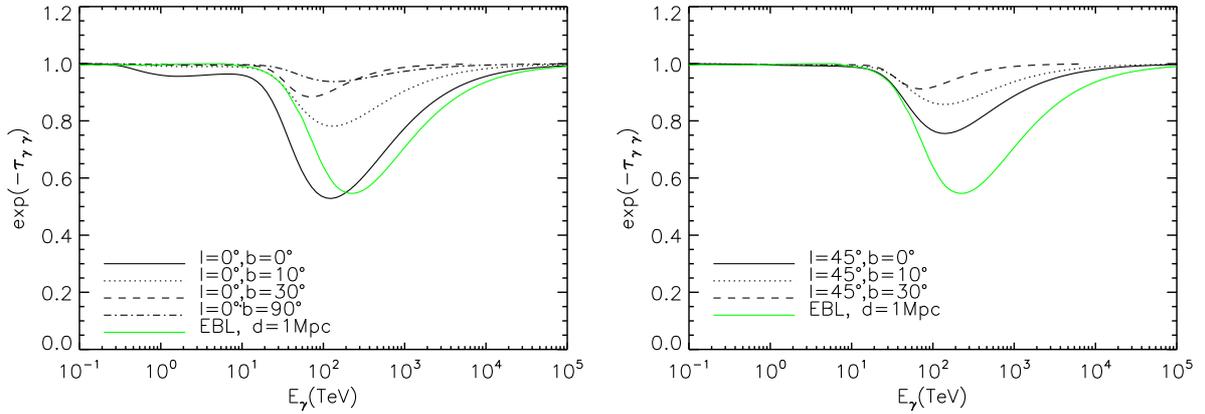}
\caption{Left: the transmission curve for gamma-ray extragalactic sources
  located at galactic coordinates:
$(l=0^{\circ}, b=0^{\circ})$ - solid line, 
$(l=0^{\circ}, b=10^{\circ})$ - dotted line,  
$(l=0^{\circ}, b=30^{\circ})$ - dashed line and
$(l=0^{\circ}, b=90^{\circ})$ - dashed-dotted line. Right: the transmission
curve for gamma-ray extragalactic sources located at galactic coordinates: 
$(l=45^{\circ}, b=0^{\circ})$ - solid line, 
$(l=45^{\circ}, b=10^{\circ})$ - dotted line and 
$(l=45^{\circ}, b=30^{\circ})$ -dashed line. For these curves only the opacity 
due to pair production resulting from interactions between gamma-rays and 
photons of the ISRF in the Milky Way is considered. The contribution due to 
the CMB and EBL is not included, as this would depend on the distance to the 
extragalactic sources. For comparison the green curves show the effect on the
transmission due only to the EBL light for a source located at a fiducial
distance of 1\,Mpc.}
\label{fig:tau_extragalactic}
\end{figure*}

\subsection{Gamma-ray opacity} 

The same ISRF that dominate the production
of Inverse Compton radiation by CR electrons are also responsible for
an attenuation of gamma-rays. This attenuation arises through pair production 
resulting from interactions between the gamma-rays and photons of the ISRF. 
The energy dependence of the pair production cross-sections have been 
given by Gould \& Schr{\'e}der (1966). As a rough rule of thumb, the opacity to 
gamma-rays of energy $E_{\gamma}$ interacting with photons of wavelength 
$\lambda$ peaks for ($E_{\gamma}$/TeV)/($\lambda/\mu$m) $\sim1$, with a sharp cut 
off towards shorter wavelengths. Thus, while the CMB controls the visibility 
of sources of gamma-rays with $E_{\gamma}$ around 1000\,TeV, radiation fields 
in the range $0.1-100$\,$\mu$m control the visibility of sources in the 
$0.1-100$\,TeV range which is the typical energy range over which ground-based
Cherenkov imaging telescopes such as HESS, Veritas, and in the future, CTA,
operate. In the following we show how the solution we have obtained
for radiation fields in the Milky Way in this wavelength range affect the
visibility of galactic and extragalactic gamma-ray sources as a function of
their galactic coordinates and (in the case of galactic sources) their
distance from the Sun. The $\gamma$-$\gamma$ opacity was calculated by 
performing the line-of-sight 
integral of the product of the ISRF energy density and the pair production 
cross section.

Fig.~\ref{fig:tau_galactic} shows the transmission curve (i.e. the fraction of 
photons reaching the Sun as a function of photon energy) for gamma-ray sources 
located at the Galactic Centre, and in the plane of the galaxy at a location 
8\,kpc from the Galactic Centre in the opposite direction to the Sun.
Two main peaks in opacity are seen at ca. 2000 and 100\,TeV, respectively
corresponding to the CMB and the far-IR peak in the ISRF. The opacity peak
associated with the CMB is deeper than that associated with the
ISRF, since, though the energy density in the far-IR is comparable to the 
0.26\,eV/${\rm cm}^3$ of the CMB over the paths through the plane of the 
galaxy, the number density of the CMB photons is higher by about an order of 
magnitude. For the same reason, the UV/optical/NIR photons in the disk have a 
negligible effect on source visibility. The only exception is a ledge in the 
transmission curve from ca. 2 to 10\,TeV, which is due to the line of sight 
passing  through the intense radiation NIR fields produced by the bulge in 
the region  of the Galactic Centre. As might be expected due to the 
cylindrical symmetry 
of the calculation of the ISRF, the visibility of the source at the Galactic 
Centre is, at all energies, higher by a constant factor corresponding to factor 
of 2 in optical depth, compared to the source 8\,kpc beyond the Galactic 
Centre. 

An interesting feature of the plots is that already at energies of 
50\,TeV, a significant reduction of visibility of sources at the Galactic 
Centre is expected. This is physically attributable to the predicted sharp 
up-turn in radiation energy density at 20\,${\mu}$m (see - top left panel of 
Fig.~\ref{fig:rf_dust_profiles}) from diffuse dust in the disk in the 
vicinity of the Galactic Centre heated by optical photons from the inner cusp 
of the bulge close to the Galactic Centre. This illustrates that even small 
efficiencies of conversion of optical to infrared photons through absorption 
by dust can have a disproportionately strong effect on the pair production 
opacity of gamma-rays. The extreme steepness of the decrease in transparency 
between 20 and 100\,TeV is due to a combination of two factors: Firstly, the pair 
production opacity is due to photons on the Wien side of the FIR dust emission 
peak, so that in progressing to slightly higher gamma-ray energies brings a 
relatively large increment in energy density of longer wavelength infrared 
photons into play. And secondly, this effect is compounded by the even larger
number densities of the longer wavelength infrared photons involved. We note 
that, since the infrared radiation fields on the Wien side of the dust 
emission spectrum have a strong component of locally heated dust, the depth 
and steepness of the transmission curve between 20 and 100\,TeV is  
dependent on the presence  and amount of dust in the inner few tens of parsec 
of the disk of the Milky Way. This makes the visibility of pevatrons at, or 
seen through the Galactic Centre very dependent on complex physical processes 
controlling dust abundance in this region.

\begin{figure}
\centering
\includegraphics[scale=0.5]{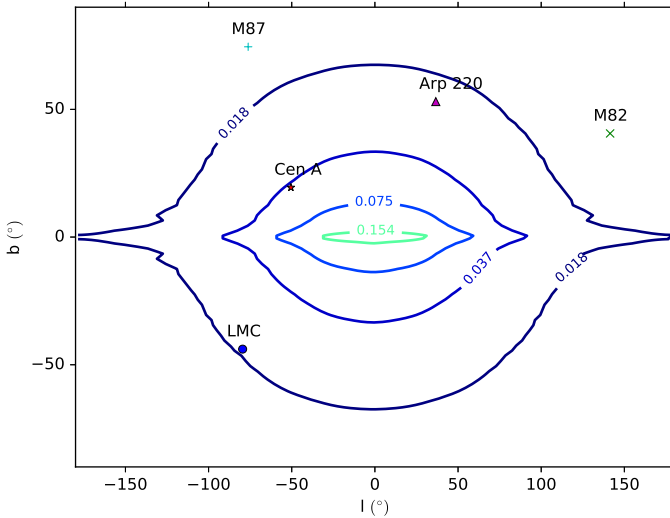}
\caption{Contours of opacity at 40\,TeV due to pair production in the predicted
  ISRF of the Milky Way for extragalactic sources as a function of Galactic 
longitude $l$ and latitude $b$. The positions of prominent nearby very high 
energy sources are given on the plot.}
\label{fig:opacity_contours}
\end{figure}

To examine the implications of our solution for the ISRF for the visibility of
extragalactic sources we have plotted in Fig.~\ref{fig:tau_extragalactic} 
transmission curves for
external sources at selected galactic coordinates. To isolate the effect of the
ISRF, these curves do not take into account the attenuation due to the CMB or
the extragalactic background (EBL). For the purpose of comparison a further 
transmission curve showing the pair
production absorption due to the EBL alone is 
plotted for a source located at a fiducial distance of 1\,Mpc. For this we 
used the EBL derived by Franceschini et al. (2008). As expected, the position of the
peak in absorption due to the EBL is at higher gamma-ray energies than that due
to the ISRF of the Milky Way. This is because the SED of the EBL peaks at 
longer submm wavelengths, due to the EBL light being dominated by emission from
cosmologically distant dusty sources. Nevertheless, even in the energy range
up to 100\,TeV, the effect of the ISRF on source visibity only approaches
that of the EBL for sources in the Local Group (at distances scales of
ca. 1Mpc) at low galactic latitudes. In the 
0.1 to 20\,TeV energy range, in which extragalactic sources have been 
commonly measured, the attenuation due to the ISRF is negligible, except
for lines of sight into the inner galaxy, which in any case are avoided for
extragalactic studies. A contour map in galactic coordinates 
of the 40\,TeV opacity towards extragalactic sources due to pair production 
ISRF is given in Fig.~\ref{fig:opacity_contours}. We note that on this plot 
the contours are
perfectly symmetrical between North and South, as we have taken the 
Sun to be located exactly in the plane of the Milky Way. In reality the Sun is
displaced about ca. $15-20$\,pc North of the plane (e.g Cohen 1995, Reed 2006,
Joshi 2007), so the Northern sky will, at
a certain level, be more transparent than the Southern sky. However, this 
introduces an asymmetry only of the order of $1\%$ in the opacities between 
North and South.

\section{Influence of the modelled ISRF on predicted VHE Emission from
accelerators of CR electrons}
\label{sec:accelerators}

\begin{figure*}
\centering
\includegraphics[scale=0.9]{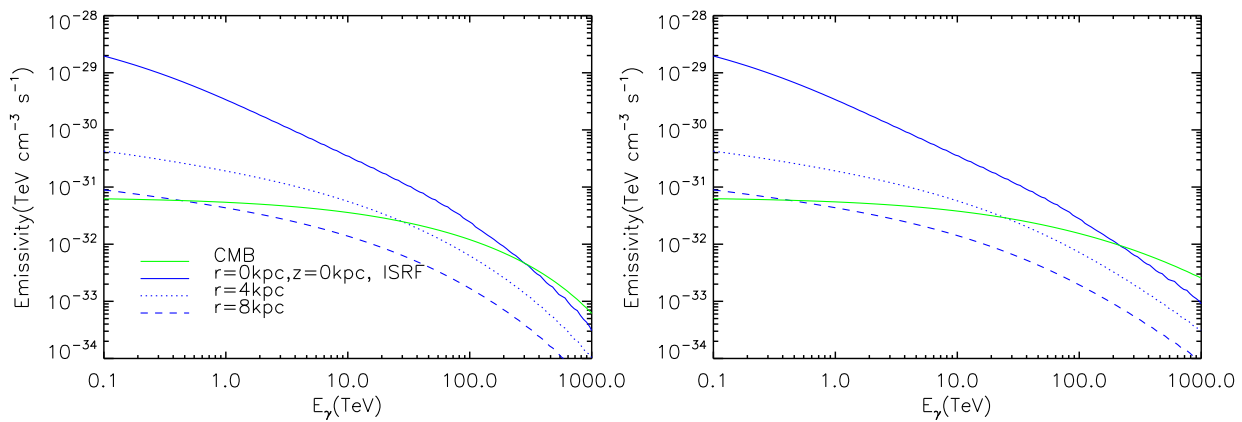}
\caption{Predicted VHE IC emission from scattering of photons from the ISRF,
  compared to that from the CMB at different position in the galaxy. The
  electron spectrum has a power law index of -3 and either an energy cut-off 
$E_{\rm b}=1000$\,TeV (LH panel) or no cut-off energy ( RH panel).} 
\label{fig:accelerators} 
\end{figure*}

Inverse-Compton
scattering by CR electrons of either ISRF or CMB photons, 
is generally not invoked to account for a major fraction 
of the VHE gamma-ray emission ($>\sim 1$\,TeV)
from the diffuse ISM, due to the very high radiation
losses of the electrons as they propagate away from 
their sources. For this reason, we have
considered only electron energy spectra with cut offs
of 1TeV and below when considering the IC emission 
from the diffuse ISM in Sect.~\ref{subsec:IC}.
While the IC mechanism is commonly considered
when accounting for the observed VHE emission
in putative sources of CR, generally only scattering 
of CMB electrons has been considered. The reason for this is
the predicted suppression of the scattering cross section
of ISRF photons compared to that for the CMB photons, 
due to the Klein Nishina effect. 
Here we show that in fact the ISRF has a strong effect
on both the hardness and amplitude
of the inferred energy distribution of CR electrons
in sources in the galactic plane, especially
within the solar circle.

This is illustrated by Fig.~\ref{fig:accelerators}, which shows 
the predicted VHE IC emission from
scattering of photons from the ISRF, compared to that from
the CMB at different
positions in the galaxy. The same electron spectrum
with a power law index of -3 as considered in Fig.~\ref{fig:sed_ic}
has been used for these predictions,
but now with either no energy cut off at all (RH panel) or
with a cut off at 1000\,TeV (LH panel). As expected,
the spectral slope of the emission from the ISRF
steepens at high energy compared to that from the CMB.
However, 
the curves for IC from the ISRF both with and without
the cut off in intrinsic electron energy spectrum only cross the
curve for IC from the CMB at about 0.4, 30, and $300$\,TeV
for sources in the plane situated at galactocentric radii of
respectively 8, 4 and 0\,kpc.

This shows that in general,
the component of VHE IC emission from the ISRF
will be important, and in many cases dominant, for sources within
the solar circle. IC from the ISRF should therefore be considered
when discussing leptonic hypotheses for the CRs giving rise to observed
VHE gamma-ray emission from such sources. Most particularly,
we note that, due to the Klein-Nischina effect, a harder energy
spectrum of CR electrons will be inferred 
under this hypothesis compared to the case that only IC scattering
off the CMB is considered, for sources located in the inner galaxy 
where the ISRF component dominates out to a few tens of TeV in
photon energy. In general, this hardening will have to be considered,
on a source-by-source basis,
in conjunction with independent considerations influencing the hardness of the
energy spectum of electrons in the sources, when evaluating the
leptonic hypothesis. This includes consideration of the predicted
intrinsic hardness of the electron energy spectrum at injection
for putative acceleration mechanisms, and/or the effect 
of radiation losses in the sources from the combination of
synchrotron and IC losses.

\section{Summary}
\label{sec:summary}

We use the formalism of the radiation transfer model of 
Popescu et al. (2011) and a new methodology that deals with the inner view of a
galaxy and with the lack of direct observational constraints in the UV-optical
regime to derive an axisymmetric solution for the ISRF
of the Milky Way in direct and dust-reradiated starlight
from  912\,\AA to 1\,mm, over the volume within 24\,kpc in galactocentric
radius and $\pm 10$\,kpc in height above the plane. The model of the Milky Way 
was obtained by fitting the submm, FIR, MIR  and NIR observed maps as seen from the 
position of the Sun. 

The calculations were made using a modified version of the 2D ray-tracing 
radiative transfer code of Kylafis \& Bahcall (1987), and the 3D
ray-tracing RT code DART-Ray (Natale et al. 2014, 2015). Both codes 
include anisotropic scattering, and the dust model from Weingartner \& Draine 
(2001) and Draine \& Li (2007), incorporating a mixture of silicates, graphites
and PAH molecules. The model for the dust emission incorporates a full
calculation of the stochastic heating of small grains and PAH molecules. 
The geometrical model consists of a large-scale distribution of 
diffuse dust and stars, as well as of
a clumpy component physically associated with the star-forming
complexes. The large-scale distribution of stars consists of a thick and a thin
stellar disk, a inner thin stellar disk and a bulge, all seen through a common
distribution of dust. The diffuse dust is distributed into a thick and thin
dust disk, respectively, on account of the different thickness of these 
structures.

We find that our model is able to account for the overall observed stellar 
and dust emission SED of the Milky Way (as seen from the position of the Sun), 
as derived from IRAS, COBE and Planck maps at 1.2, 2.2, 3.5, 4.9, 24, 60, 
100, 140, 240, 350, 550 and the 850${\mu}$m. At 1.2\,${\mu}$m the model
underestimates the emission by $27\%$, possibly indicating higher dust
efficiencies in the submm than the WD01 model used in this paper or more
complex geometries in the inner region due to the boxy-peanut shape of the
bar/bulge structure.
The agreement in the MIR, where the dust emission is 
predominantly powered by UV, means that the model prediction 
for the illumination of grains by UV light is good, both in the diffuse
ISM and locally in the star-formation regions. This is the most direct
constraint of the global distribution of UV light in the Milky Way,
where direct measurement of UV light can only be performed for nearby
stars. The correct prediction of the 
colour between the peak of the SED
in the FIR and the MIR shows that both the $SFR$ and the probability of escape
of non-ionizing UV light from star-formation regions in the diffuse ISM are
also correctly modelled. Our model of the Milky Way is thereby the first to 
consider all accessible emission components in a self consistent radiation
transfer treatment embodying absorption, anisotropic scattering,
and non-equilibrium emission from transiently heated grains, as well
as local absorption from star forming regions.

Comparison with the previously frequently used model GALPROP shows that
the radiation fields calculated by GALPROP systematically differ from those 
predicted by our model. We ascribe the differences to the fact that 
GALPROP was not optimised to fit the all-sky multiwavelengths images.

We describe and account for the spatial and spectral variations in
the derived ISRF as a function of galactocentric position, and
explore the imprint of these variations on the amplitude and
spectrum of the component of gamma-ray emission due to inverse-Compton
 emission from diffusely distributed CR electrons in the energy range from 1
 MeV to TeV energies, using canonical reference spectra
of electrons with a variety of cut-off energies.  In general, IC from
the UV-FIR component of the ISRF dominates IC from the
cosmic microwave background (CMB)
for IC emission over the volume out to galactocentric radii of around
10\,kpc, and up to heights of 10\,kpc, with only a moderate energy
dependence of this volume of influence on the energy of the gamma-rays.
We also show that the FIR component of the ISRF, as predicted by our model,
can strongly influence the derived amplitude and slope of the very high energy 
(VHE) IC emission above 1\,TeV photon energy 
from discrete sources of CR electrons, compared to solutions only
considering the component of IC from the CMB. This is despite the suppression
of the cross-section for IC scattering of the ISRF at high energy
due to the Klein-Nishina effect. 
Finally, we compute the pair-production opacity at VHE energies up to 100\,TeV
for gamma-rays from sources located at various positions in our galaxy, 
as well as for extragalactic sources, giving a map
of variation of opacity over the sky for the latter at a
photon energy of 40\,TeV. 

The solutions for the UV/optical/FIR/submm radiation fields are available in 
electronic format for use in the analysis of gamma-ray emission in our Galaxy. 
 To understand how they can be used, we need to draw attention to the
  main limitations and applicability of our model, as detailed in
  Sect.~\ref{subsec:limitations}. First of all this paper
  presents the axis-symmetric solution for the radiation fields of a Milky Way
  having a WD01 dust type. If the submm grain efficiencies were higher than 
  the predictions of the WD01 model, then this 
  would be equivalent with an overestimation of the optical depth of the
  galaxy for a model that assumes a WD01 dust. Because the
  way our model was constructed, the solutions in the UV/optical
  (B,V,I) and MIR/FIR/submm are relatively robust against the choice of dust 
  model, but the NIR solutions could be strongly altered should the dust model 
  be different.

Because of the axial symmetry,
  the model presented in this paper cannot predict the variations in the
  diffuse ISRF between the arm and inter-arm regions, nor the variations due to
  the boxy-peanut shape of the bulge/bar structure in the inner disk. From the
  point of view of gamma-ray astronomy this means that whereas the model can
  be used to predict the inverse Compton emission from the general population
  of cosmic rays interacting with the ISRF, it can only be used to place lower
  limits to the component of IC of any localised sources of cosmic rays
  geometrically associated with star-forming regions. In future work we will 
use our 3D code DART-Ray to address this 
problem and to provide non-axisymmetric solutions for the radiation fields in
the Milky Way.

  We also draw attention
  here that our model does not include any component of ISRF from a possible
  large scale emission component from a galactic halo. If such a halo would be
  shown to exist, it would increase the predictions for the pair-production
  opacity of very high energy gamma-rays beyond the predictions given in this
  paper.

\section*{Acknowledgements}
We would like to thank an anonymous referee for a very useful and
  constructive report which helped improve the paper.
Cristina C Popescu and Giovanni Natale acknowledge support from the
Leverhulme Trust Research Project Grant RPG-2013-418. The development of the
radiative transfer code DART-Ray was supported by the UK Science and 
Technology Facilities Council (STFC; grant ST/J001341/1).

This research has made use of the NASA/ IPAC Infrared Science Archive, which is
operated by the Jet Propulsion Laboratory, California Institute of Technology,
under contract with the National Aeronautics and Space Administration. 
The research has also made use of the Centre d'Analyse de Donn\'ees Etendues
(CADE; Analysis Center for extended data). Planck data has been used
in this paper. Planck (http://www.esa.int/Planck) is a project of the
European Space Agency (ESA) with instruments provided by two
scientific consortia funded by ESA member states (in particular the
lead countries France and Italy), with contributions from NASA (USA)
and telescope reflectors provided by a collaboration between ESA and a
scientific consortium led and funded by Denmark.

We acknowledge use of data on the radiation fields from GALPROP 
(http://galprop.stanford.edu). GALPROP development is supported through NASA 
grants and by the Max Planck Society.

\end{document}